\documentclass[3p,times,twocolumn]{elsarticle}

\journal{Expert Systems with Applications}
\usepackage{float}
\usepackage{verbatim}
\usepackage{subcaption}
\usepackage{placeins}
\usepackage{algorithm}
\usepackage{algorithmic}
\usepackage{amsmath}
\usepackage{hyperref}

\floatstyle{plaintop}
\restylefloat{figure}
\restylefloat{table}

\begin{document}

\begin{frontmatter}

\title{From Experimental Limits to Physical Insight:
A Retrieval-Augmented Multi-Agent Framework for Interpreting Searches Beyond the Standard Model}

\author[label1,label2]{Altan Cakir\corref{cor1}}
\ead{cakirmua@itu.edu.tr, cakir@cern.ch}

\author[label2]{Ayca Yerlikaya}

\cortext[cor1]{Corresponding author}

\address[label1]{Istanbul Technical University, Department of Data Science and Analytics, Istanbul, Turkey}

\address[label2]{Istanbul Technical University, Department of Physics Engineering, Istanbul, Turkey}

\begin{abstract}
Modern searches for physics beyond the Standard Model produce rapidly expanding literature containing heterogeneous information, including textual analyses, numerical datasets, and graphical exclusion limits. Integrating these distributed sources remains a time-consuming and manual process for physicists.

We present HEP-CoPilot, a retrieval-augmented multi-agent AI framework for the exploration and interpretation of high-energy physics literature. The system unifies textual information from publications, structured experimental data from HEPData, and reconstructed physics plots within a multimodal retrieval and reasoning architecture. By combining retrieval-augmented language models with coordinated agent workflows, it enables evidence-grounded reasoning over experimental analyses and structured interpretation of collider results.

We evaluate the framework on recent CMS searches for physics beyond the Standard Model. Case studies show that HEP-CoPilot can retrieve relevant measurements, reconstruct exclusion limits directly from HEPData records, and perform cross-paper comparisons of experimental constraints. This enables consistent, physics-aware comparison across analyses without manual data integration.

These results demonstrate that retrieval-augmented AI systems can function as scientific co-pilots for particle physics, facilitating navigation of complex literature, structuring heterogeneous evidence, and accelerating the interpretation pipeline for new physics searches.
\end{abstract}


\end{frontmatter}

\section{Introduction}

The search for physics beyond the Standard Model (BSM) is one of the central goals of contemporary high-energy physics. Experiments at the Large Hadron Collider (LHC), including those conducted by the CMS and ATLAS collaborations, have produced an unprecedented volume of results constraining possible extensions of the Standard Model. These studies probe a wide variety of theoretical scenarios, including supersymmetry, long-lived particles, heavy stable charged particles, and other exotic phenomena predicted by many BSM frameworks. Each experimental analysis typically involves complex workflows including event reconstruction, background modeling, statistical inference, and interpretation of the results in terms of theoretical models. The outcomes of these analyses are commonly presented through exclusion limits, cross-section constraints, and parameter-space interpretations derived from large-scale experimental datasets.

Despite the richness of these results, extracting actionable scientific insight from the growing body of literature remains a significant challenge. Information relevant to interpreting experimental limits is often distributed across multiple modalities, including textual descriptions, tables of numerical constraints, plots of exclusion contours, and externally hosted datasets such as those provided by HEPData. As a consequence, physicists attempting to compare results across analyses or reinterpret limits in the context of new models must manually inspect numerous publications and datasets. This process is time-consuming, error-prone, and increasingly difficult as the volume of published analyses continues to grow.

Recent advances in large language models (LLMs) and retrieval-augmented generation (RAG) systems offer new opportunities to support scientific discovery by assisting researchers in navigating complex knowledge bases. In several domains, AI-driven systems have demonstrated the ability to extract structured knowledge from scientific documents and synthesize information across heterogeneous sources. However, most existing approaches focus primarily on textual reasoning and lack the capability to integrate scientific plots, numerical datasets, and domain-specific reasoning tasks that are essential in high-energy physics research.

In particular, interpreting experimental searches for BSM physics requires the integration of heterogeneous information sources and domain-aware reasoning. For example, reconstructing exclusion limits from HEPData tables, comparing constraints across different analyses, or interpreting the implications of a new search result for a specific theoretical scenario requires both structured data retrieval and contextual reasoning grounded in particle physics knowledge. Conventional document retrieval systems are not designed to perform such multi-modal scientific reasoning tasks.

To address these challenges, we introduce \textbf{HEP-CoPilot}, a multi-agent AI framework designed to support the exploration and interpretation of high-energy physics literature. The proposed system combines retrieval-augmented language models with specialized agents capable of processing different types of scientific information, including textual content, numerical tables, and experimental plots. Through a coordinated reasoning pipeline, the framework enables automated retrieval of relevant experimental results, reconstruction of physics plots from structured datasets, and cross-paper comparisons of experimental constraints.

The architecture of HEP-CoPilot is designed to mimic aspects of collaborative scientific reasoning. A mission-control component first interprets the user's query and determines the appropriate analytical workflow. Specialized agents are then invoked to retrieve evidence, reconstruct experimental visualizations, and perform domain-aware reasoning tasks. Finally, a synthesis component integrates the results into a coherent explanation grounded in the underlying experimental data.

The capabilities of the proposed framework are evaluated through a series of case studies based on recent searches for physics beyond the Standard Model. These experiments illustrate how the system can retrieve experimental measurements, reconstruct exclusion limits from structured HEPData records, and support comparative analysis of experimental constraints across multiple collider searches. The results indicate that AI-assisted systems can facilitate the exploration of complex experimental literature and support physicists in interpreting results reported across heterogeneous sources of scientific evidence.

The main contributions of this work can be summarized as follows:

\begin{itemize}

\item \textbf{Domain-aware AI framework for particle physics literature analysis.}  
A multi-agent system, \textbf{HEP-CoPilot}, is introduced to support interpretation of experimental searches in high-energy physics literature by coordinating retrieval, reasoning, and evidence synthesis across heterogeneous scientific sources.

\item \textbf{Multimodal retrieval over scientific evidence.}  
A unified retrieval architecture is developed that integrates textual descriptions from research papers, structured numerical datasets from HEPData, and graphical information from experimental plots within a single evidence retrieval pipeline.

\item \textbf{Reconstruction and comparison of collider exclusion limits from structured data.}  
The framework enables automated reconstruction of experimental exclusion limits directly from HEPData numerical records, allowing quantitative comparison of experimental constraints across multiple analyses and supporting physics-aware reasoning over collider search results.

\end{itemize}

Unlike conventional literature retrieval systems that rely primarily on textual summaries, the proposed framework operates directly on the numerical measurements underlying experimental plots. By reconstructing exclusion limits from structured HEPData datasets, the system enables quantitative analysis and comparison of collider search constraints reported across multiple publications.

To the best of our knowledge, this work represents one of the early attempts to develop an AI-assisted framework for interpreting experimental constraints in high-energy physics literature using structured experimental data.

\section{Related Work}

Recent advances in large language models (LLMs) and agent-based artificial intelligence systems have significantly expanded the capabilities of automated knowledge synthesis and scientific reasoning. In complex research domains such as high-energy physics (HEP), knowledge is distributed across heterogeneous sources including technical publications, structured experimental datasets, and graphical representations such as exclusion plots. Consequently, AI systems capable of integrating multiple modalities of scientific information are becoming increasingly important for supporting literature-driven scientific workflows.

This section reviews prior research across four relevant areas: (i) AI systems for scientific discovery, (ii) retrieval-augmented generation for scientific literature analysis, (iii) multimodal understanding of scientific documents, and (iv) applications of artificial intelligence in particle physics. Finally, we discuss recent approaches for extracting structured information from scientific figures and tables and position our work within this emerging research landscape.

\subsection{AI Systems for Scientific Discovery}

A growing body of research explores the use of large language models to support or partially automate the scientific discovery process. In particular, multi-agent LLM systems have emerged as a promising paradigm for decomposing complex reasoning tasks into coordinated workflows executed by specialized agents.

Lu et al.~\cite{lu2024ai_scientist} introduced the \emph{AI Scientist} framework, which demonstrates an automated pipeline capable of generating research hypotheses, proposing experiments, and producing scientific manuscripts. Similarly, Google's \emph{AI co-scientist} system illustrates how collaborative LLM agents can iteratively propose, critique, and refine hypotheses in order to accelerate discovery cycles~\cite{google_co_scientist}. Multi-agent debate frameworks have further shown that structured interactions between multiple models can improve reasoning quality and scientific review processes by enabling models to critique and refine candidate explanations~\cite{xu2023large_scale_review}.

These studies highlight the potential of agent-based architectures for supporting scientific reasoning. However, most existing systems focus on general scientific tasks or domains such as biology and chemistry, while relatively little work addresses the specific challenges associated with interpreting experimental results in high-energy physics.

\subsection{Retrieval-Augmented Generation for Scientific Literature}

Retrieval-Augmented Generation (RAG) has become a key technique for improving the factual reliability of LLM outputs when operating over large knowledge bases. By retrieving relevant documents and conditioning the generation process on external evidence, RAG systems enable language models to ground their responses in verifiable sources.

Lewis et al.~\cite{lewis2020rag} introduced the foundational RAG architecture for knowledge-intensive natural language processing tasks, demonstrating how neural retrieval models combined with generative language models can significantly improve question answering performance. Building upon this idea, Lála et al.~\cite{lala2023paperqa} proposed \emph{PaperQA}, a system designed to answer complex scientific questions by retrieving and synthesizing information from full-text research papers. More recent studies further investigate the use of LLM-based retrieval pipelines for scientific literature exploration and knowledge synthesis across multiple domains~\cite{ramprasad2025}.

Despite these advances, most RAG-based approaches remain largely text-centric and do not adequately capture the multimodal structure of scientific publications.

\subsection{Multimodal Understanding of Scientific Documents}

Scientific publications are inherently multimodal, combining textual explanations with tables, figures, equations, and supplementary datasets. Automated reasoning over such documents therefore requires models capable of integrating heterogeneous information modalities.

Recent benchmark datasets such as SPIQA~\cite{pramanick2024spiqa} and SciTabQA~\cite{ghosh2023scitabqa} have been developed to evaluate multimodal question answering over scientific documents. These tasks require models to interpret textual descriptions together with tabular data and graphical elements. However, existing studies indicate that even advanced vision-language models still struggle with tasks involving quantitative reasoning, numerical comparison, or interpretation of scientific plots.

This limitation is particularly significant in high-energy physics, where exclusion plots and constraint curves serve as the primary mechanism for communicating experimental limits.

\subsection{Artificial Intelligence in Particle Physics}

Machine learning techniques have played an important role in particle physics for several decades. Early studies explored the use of neural networks for pattern recognition in detector signals~\cite{denby1988}. More recently, deep learning approaches have been widely adopted for event classification, detector reconstruction, and anomaly detection in collider experiments~\cite{radovic2018ml}. Reviews of modern machine learning applications at the Large Hadron Collider highlight the increasing integration of AI across multiple stages of the experimental pipeline, ranging from trigger systems to high-level physics analysis~\cite{duarte2024}.

In parallel, several domain-adapted language models have been proposed to better capture the vocabulary and structure of physics literature. Models such as Astro-HEP-BERT~\cite{simons2024astrohepbert}, AstroLLaMA~\cite{nguyen2023astrollama}, and PhysBERT~\cite{hellert2024physbert} demonstrate improved performance on tasks such as document retrieval and concept linking within scientific corpora. Nevertheless, these models primarily focus on textual representations and do not provide integrated mechanisms for reasoning across textual descriptions, numerical datasets, and experimental plots.

\subsection{Extraction of Scientific Data from Figures and Tables}

Another line of research focuses on the automated extraction of structured information from scientific figures and tables. Recent approaches leverage vision-language models and chain-of-thought prompting techniques to digitize data from plots or recover numerical values from complex graphical representations.

For example, PlotExtract~\cite{polak2024plotextract} proposes a reasoning-based method for recovering data points from scientific plots, enabling subsequent quantitative analysis. While such tools represent important progress toward automated scientific data extraction, they are typically designed as standalone systems and do not provide an integrated reasoning framework capable of combining extracted numerical data with textual evidence and domain knowledge.

\subsection{Positioning of This Work}

Although previous research demonstrates the potential of multi-agent LLM systems, retrieval-augmented generation pipelines, and domain-adapted language models for scientific literature analysis, existing approaches rarely address the specific challenges posed by high-energy physics publications. In particular, interpreting experimental searches for physics beyond the Standard Model requires integrating textual descriptions, numerical constraints, and graphical representations of exclusion limits.

To address this gap, we propose \textbf{HEP-CoPilot}, a retrieval-augmented multi-agent framework designed to support literature-driven reasoning over experimental particle physics results. The system integrates multimodal scientific evidence from research papers and structured experimental datasets provided by HEPData, enabling automated reconstruction and comparison of experimental exclusion limits across multiple analyses.

The framework is evaluated using recent CMS analyses searching for new physics signatures in proton-proton collisions at $\sqrt{s}=13$ TeV~\cite{CMS_HSCP_2024,CMS_LLP_2024,CMS_STOP_2025}. Unlike general-purpose scientific question answering systems, the proposed framework is specifically designed to support physics-aware reasoning over experimental collider search results.

\section{Methodology}

Traditional retrieval-augmented generation (RAG) systems are primarily designed for textual question answering and lack the capability to reason over heterogeneous scientific evidence. In contrast, high-energy physics analyses involve complex experimental measurements distributed across multiple modalities, including textual descriptions, structured datasets, mathematical expressions, and graphical representations such as exclusion plots.

To address this challenge, we propose \textbf{HEP-CoPilot}, a multimodal scientific reasoning framework designed to assist physicists in interpreting experimental particle physics results. The proposed system integrates retrieval-augmented evidence collection with a coordinated multi-agent reasoning architecture that enables structured analysis of experimental constraints reported in scientific publications and associated datasets.

The architecture combines multimodal knowledge representation, scientific document retrieval, and agent-based reasoning to generate evidence-grounded explanations for physics queries. The overall system architecture is illustrated in Figure~\ref{fig:system_architecture}.

\subsection{System Architecture}

HEP-CoPilot processes a user query through a multi-stage pipeline that integrates query interpretation, multimodal evidence retrieval, coordinated reasoning across specialized agents, and response synthesis.

Formally, the system can be expressed as a functional pipeline mapping a user query $q$ to a response $y$:

\[
y = S(P(R(q)))
\]

where $R(\cdot)$ denotes the multimodal retrieval process, $P(\cdot)$ represents the agent-based planning and reasoning pipeline, and $S(\cdot)$ is the synthesis module responsible for generating the final explanation.

\begin{figure*}[!t]
\centering
\includegraphics[width=\linewidth]{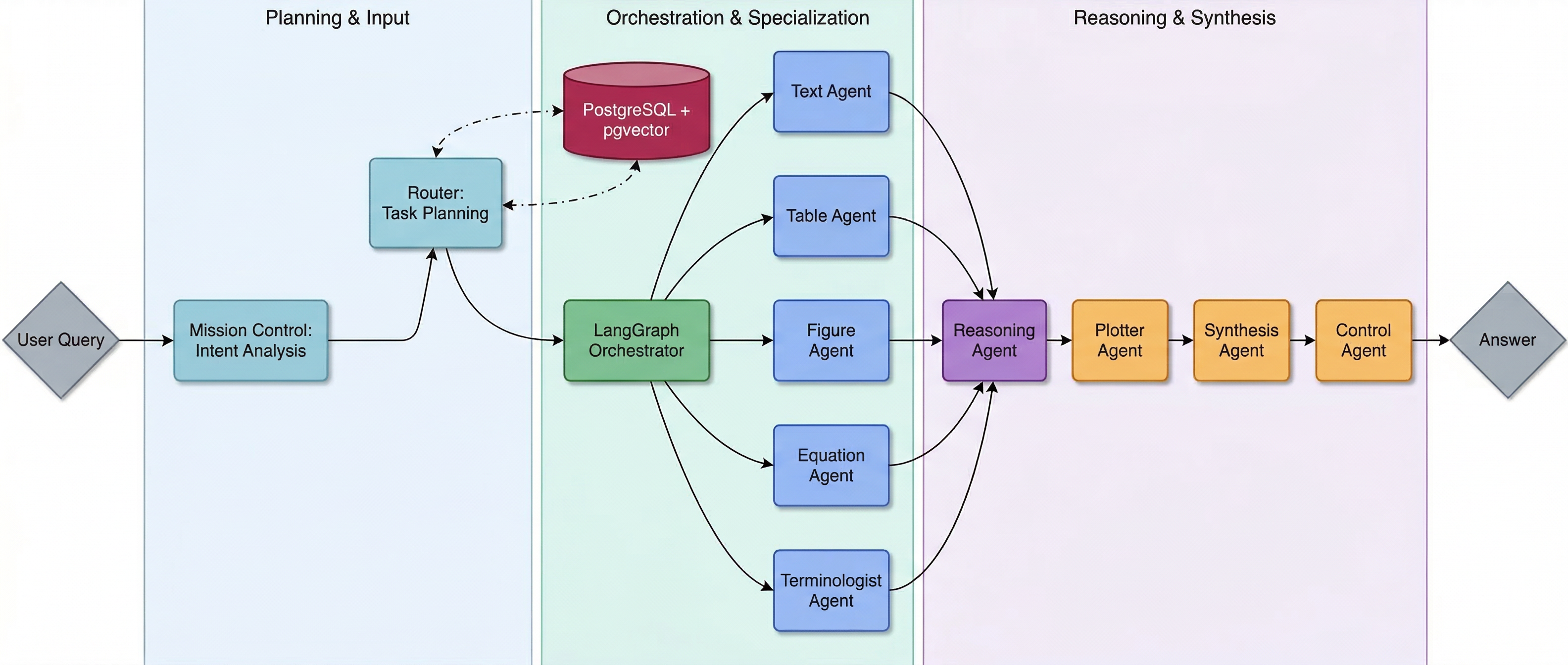}
\caption{Overall multi-agent architecture of the HEP-CoPilot framework. 
A Mission Control module first analyzes the user query and determines the analytical objective. 
A Router performs task planning and coordinates the execution graph managed by the LangGraph orchestrator. 
Specialized agents process different scientific modalities including textual content, numerical tables, figures, equations, and domain terminology. 
The outputs of these agents are integrated by a reasoning module, followed by visualization and synthesis components that generate the final evidence-grounded response.}
\label{fig:system_architecture}
\end{figure*}

The architecture is organized into three primary stages, corresponding to the workflow illustrated in Figure~\ref{fig:system_architecture}:

\begin{enumerate}
\item Query interpretation and task planning
\item Multi-agent orchestration over multimodal scientific evidence
\item Reasoning-based synthesis of experimental results
\end{enumerate}

These stages collectively enable structured reasoning over heterogeneous scientific information sources.

\subsection{Multimodal Scientific Knowledge Representation}

Scientific publications in particle physics contain information distributed across multiple modalities. Experimental analyses typically include textual descriptions of search strategies, numerical measurements reported in tables, mathematical expressions defining physical observables, and graphical representations such as exclusion limits or cross-section plots.

To enable unified retrieval across these heterogeneous sources, each scientific document $d$ is decomposed into a set of modality-specific information units:

\[
d = \{t_1, t_2, ..., t_n, g_1, g_2, ..., g_m, s_1, s_2, ..., s_k\}
\]

where

\begin{itemize}
\item $t_i$ represents textual segments extracted from research papers,
\item $g_j$ denotes graphical elements such as figures or plots,
\item $s_k$ corresponds to structured numerical entries derived from tables or external datasets such as HEPData.
\end{itemize}

Each information unit $x_i$ is encoded into a vector representation:

\[
e_i = f_{\text{embed}}(x_i)
\]

where $f_{\text{embed}}(\cdot)$ denotes the embedding function used to map scientific content into a shared semantic vector space.

The resulting embeddings are stored in a vector database implemented using PostgreSQL with the \textit{pgvector} extension. This enables efficient similarity-based retrieval across heterogeneous scientific modalities.

\subsection{Retrieval-Augmented Scientific Evidence Collection}

When a user submits a query $q$, the system retrieves relevant scientific evidence using a retrieval-augmented generation pipeline. The query is first embedded into the same semantic space used to index the knowledge base:

\[
e_q = f_{\text{embed}}(q)
\]

A similarity search is then performed within the vector database to identify the top-$k$ most relevant information units:

\[
E_q = \{x_1, x_2, ..., x_k\}
\]

The retrieved evidence set $E_q$ may include textual descriptions from scientific publications, numerical measurements from experimental datasets, or metadata describing graphical representations of experimental results.

\begin{figure}[!t]
\centering
\includegraphics[width=0.9\linewidth]{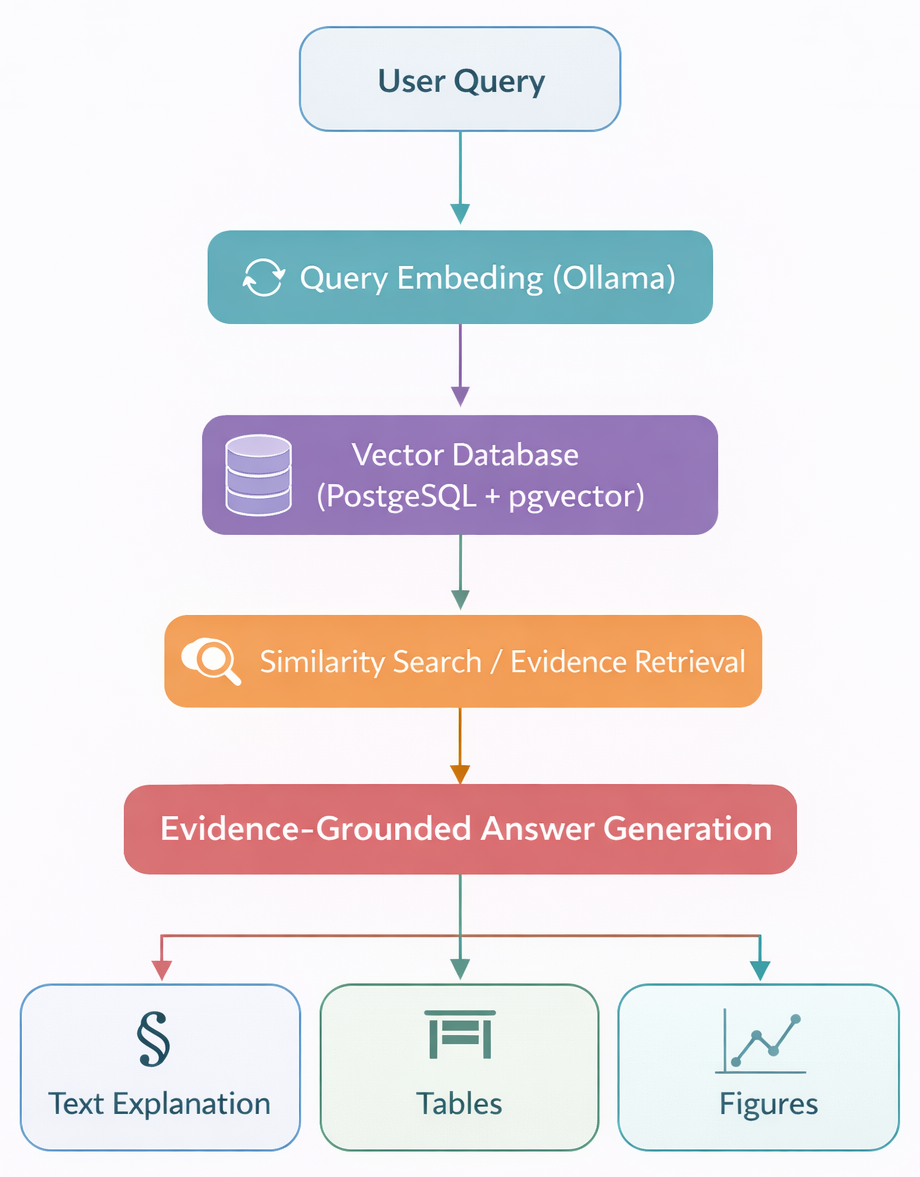}
\caption{Retrieval-augmented query processing workflow in HEP-CoPilot.
A user query is first converted into a vector representation using an embedding model. 
The query embedding is used to perform similarity search over a PostgreSQL vector database containing embedded scientific evidence. 
Retrieved evidence chunks provide contextual grounding for the language model, enabling generation of responses supported by experimental data, including textual explanations, structured tables, and reconstructed figures.}
\label{fig:rag_pipeline}
\end{figure}

This evidence set forms the contextual foundation for the downstream reasoning process executed by the agent framework. The retrieval workflow used in HEP-CoPilot is illustrated in Figure~\ref{fig:rag_pipeline}.

\subsection{Agent-Based Reasoning Architecture}

To enable structured scientific reasoning, HEP-CoPilot employs a modular multi-agent architecture implemented using a graph-based orchestration framework. In this architecture, specialized agents collaborate to process different modalities of scientific information and perform task-specific reasoning operations.

The reasoning workflow is modeled as a directed computational graph:

\[
G = (V, E)
\]

where each node $v \in V$ represents a specialized agent and edges $e \in E$ represent the flow of intermediate information between agents.

The execution of this graph is coordinated through a \textbf{LangGraph-based orchestration layer}, which dynamically manages task dependencies and information exchange between agents.

\subsection{Query Interpretation and Task Planning}

The reasoning process begins with the \textbf{Mission Control Agent}, which performs semantic analysis of the user query. This component identifies the underlying analytical objective of the query, such as retrieving experimental constraints, reconstructing exclusion limits, or comparing results across multiple collider searches.

Based on this interpretation, the \textbf{Router Agent} performs task planning by determining which specialized agents should be activated and defining the execution pathway within the reasoning graph.

This planning stage enables the system to dynamically adapt the reasoning workflow depending on the type of scientific analysis requested by the user.

\subsection{Specialized Analytical Agents}

Once the reasoning plan is established, the LangGraph orchestrator coordinates the execution of several specialized agents responsible for processing different scientific modalities.

\textbf{Text Agent.}  
Processes textual content extracted from scientific publications. The agent analyzes sections of research papers, including descriptions of experimental analyses, interpretations of results, and figure captions, and identifies statements relevant to the user query.

\textbf{Table Agent.}  
Handles numerical information reported in tables within scientific publications and experimental datasets. The agent extracts structured measurements such as cross-section limits, branching ratios, and mass exclusion values from tabulated results.

\textbf{Figure Agent.}  
Processes graphical results associated with collider experiments. The agent retrieves numerical datasets corresponding to experimental plots, primarily from repositories such as HEPData, enabling access to the measurements underlying published figures.

\textbf{Equation Agent.}  
Analyzes mathematical expressions appearing in the scientific documents. The agent extracts physical quantities and relationships defined in equations that are relevant to the interpretation of experimental results.

\textbf{Terminologist Agent.}  
Provides domain-specific interpretation of particle physics terminology. The agent ensures consistent understanding of experimental concepts and terminology across different scientific publications.

\subsection{Scientific Reasoning and Visualization}

Outputs generated by the specialized agents are integrated by the \textbf{Reasoning Agent}, which performs higher-level scientific analysis. This component combines evidence retrieved from multiple sources and conducts comparative interpretation of experimental constraints.

To support visualization of experimental results, the \textbf{Plotter Agent} reconstructs scientific plots using numerical data retrieved from experimental datasets such as HEPData. This allows the system to reproduce exclusion curves and other physics visualizations directly from structured measurements.

\begin{figure*}[!t]
\centering
\includegraphics[width=\linewidth]{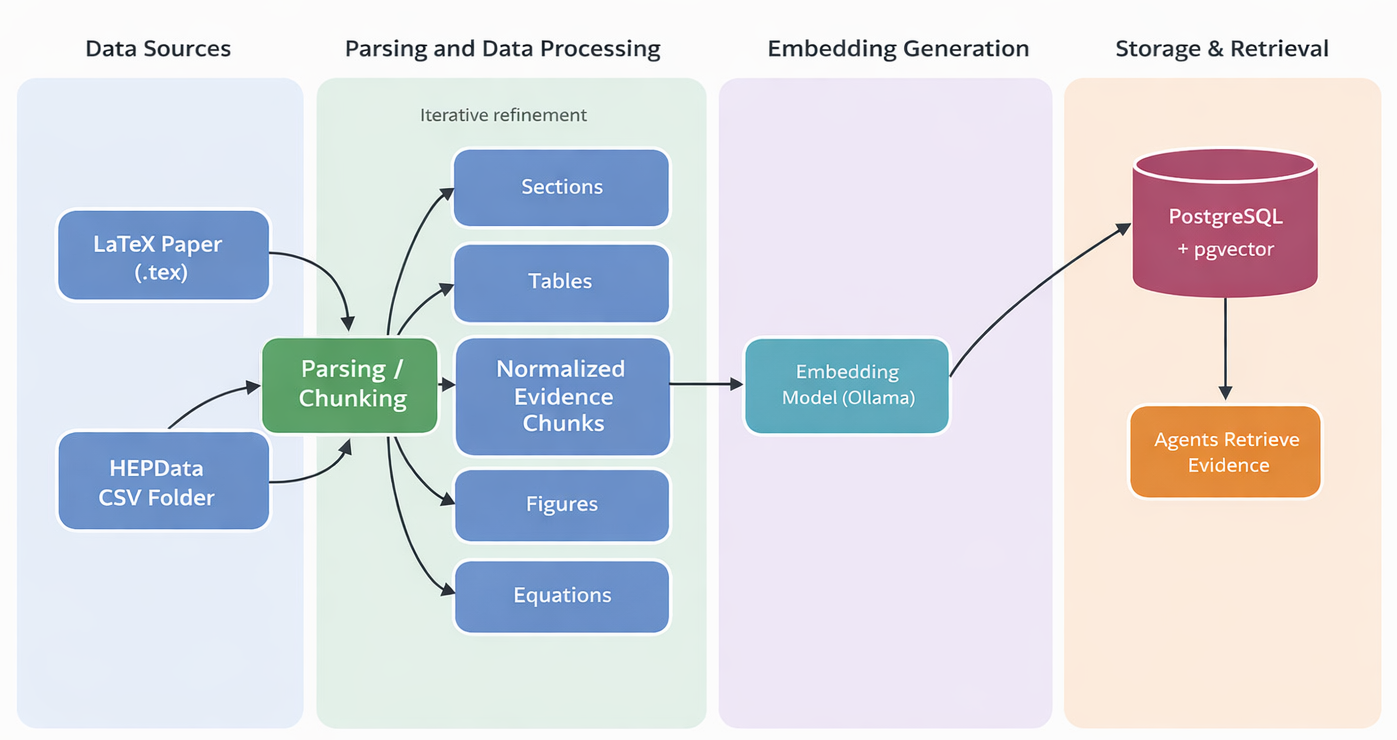}
\caption{Multimodal scientific data processing pipeline used in HEP-CoPilot.
Scientific publications written in LaTeX and structured datasets obtained from the HEPData repository are parsed into modality-specific components including textual sections, numerical tables, figures, and mathematical equations. 
These elements are normalized into evidence chunks and converted into dense vector embeddings using an embedding model accessed through the Ollama framework. 
The resulting embeddings are stored in a PostgreSQL database extended with the pgvector module, enabling efficient similarity-based retrieval of heterogeneous scientific evidence during query processing.}
\label{fig:data_pipeline}
\end{figure*}

\subsection{Evidence-Based Response Synthesis}

The final stage of the pipeline integrates all intermediate outputs into a coherent explanation addressing the user query.

The \textbf{Synthesis Agent} aggregates reasoning outputs and constructs a structured response grounded in retrieved experimental evidence. Before generating the final output, the \textbf{Control Agent} validates intermediate results to ensure consistency between generated explanations and the retrieved scientific data.

This validation step reduces the risk of hallucinated interpretations and ensures that responses remain faithful to the underlying experimental measurements.

Through this multi-agent architecture, HEP-CoPilot can retrieve heterogeneous scientific evidence, reconstruct experimental plots from numerical datasets, and perform structured reasoning over collider search results reported across multiple physics publications.

\section{Experimental Setup}

This section describes the experimental setup used to construct the multimodal knowledge base and reasoning infrastructure of the proposed HEP-CoPilot framework. The setup includes the scientific data sources used to build the knowledge repository, the preprocessing pipeline that converts heterogeneous scientific documents into structured evidence units, and the implementation details of the retrieval infrastructure.

Figure~\ref{fig:data_pipeline} illustrates the complete data processing workflow used in the system. The pipeline transforms scientific publications and associated experimental datasets into normalized evidence chunks that can be embedded, indexed, and retrieved during query processing.

\subsection{Data Sources}

The knowledge base of HEP-CoPilot is constructed using two complementary sources of scientific information commonly used in high-energy physics research: scientific publications and structured experimental datasets.

The first source consists of research papers describing experimental analyses performed by large collider experiments such as the CMS and ATLAS collaborations at the Large Hadron Collider (LHC). These publications contain detailed descriptions of experimental methods, statistical procedures, and interpretations of collider search results.

However, directly processing scientific papers in PDF format presents significant limitations for automated reasoning systems. In most cases, numerical measurements contained in tables and experimental plots are embedded in the PDF as rendered images or formatted text. As a result, extracting the underlying numerical values required for quantitative reasoning is difficult or unreliable.

To address this limitation, HEP-CoPilot processes the original \LaTeX\ source files of scientific publications whenever available. Parsing the \LaTeX\ representation allows the system to accurately identify document structure and extract semantic components such as sections, tables, equations, and figure references.

The second data source consists of structured datasets obtained from the HEPData repository. HEPData serves as a public archive for numerical data associated with particle physics publications. These datasets include the precise numerical measurements underlying experimental plots and tables, such as cross-section limits, signal efficiencies, and exclusion boundaries.

By combining textual information from the \LaTeX\ source of scientific publications with structured numerical measurements from HEPData, the system can recover both the descriptive explanations and the quantitative experimental constraints required for physics reasoning.

\subsection{Parsing and Evidence Construction}

The preprocessing stage converts raw scientific inputs into normalized evidence units suitable for semantic retrieval.

Scientific publications written in \LaTeX\ are first processed through a parsing and chunking pipeline. During this stage, the document structure is analyzed and decomposed into distinct semantic components including textual sections, tables, figures, and mathematical equations. Each extracted element represents a coherent unit of scientific information.

In parallel, experimental datasets retrieved from HEPData are ingested from CSV files and converted into structured representations. These datasets contain the numerical values underlying experimental measurements together with metadata describing the corresponding observables and binning configurations.

After extraction, all information units are transformed into a unified representation referred to as \textit{normalized evidence chunks}. These chunks represent semantically meaningful pieces of scientific knowledge that can be independently embedded and retrieved during query processing.

This normalization stage enables the system to integrate heterogeneous scientific modalities—including text, tables, figures, and numerical datasets—into a single searchable knowledge representation.

\subsection{Embedding and Vector Storage}

Once evidence chunks are generated, each chunk is converted into a dense vector representation using a transformer-based embedding model accessed through the Ollama framework. The embedding model maps each evidence unit into a high-dimensional semantic vector space that captures relationships between scientific concepts and experimental measurements.

The generated embedding vectors are stored in a PostgreSQL database extended with the \textit{pgvector} module. This vector database enables efficient similarity-based retrieval by comparing the embedding representation of a user query with stored evidence vectors.

During query processing, specialized reasoning agents retrieve the most relevant evidence chunks from the vector database. These retrieved elements provide the contextual information required for the multi-agent reasoning pipeline described in the previous section.

Through this data processing pipeline, HEP-CoPilot constructs a structured multimodal knowledge base that preserves both the textual context and the numerical measurements required for interpreting experimental particle physics results.

\section{Case Studies and Results}
Unlike conventional question answering evaluations that focus on textual responses, the experiments in this study evaluate the system’s ability to operate directly on numerical measurements underlying collider physics plots. 

By reconstructing experimental exclusion limits from HEPData records, the system enables quantitative comparison of experimental constraints across multiple analyses. This evaluation therefore tests not only textual reasoning but also the system’s ability to interpret numerical experimental results reported in particle physics publications.

To evaluate the effectiveness of the proposed HEP-CoPilot framework, we conducted a series of experiments designed to assess its ability to retrieve, interpret, and synthesize information from high-energy physics publications and associated experimental datasets. The evaluation focuses on two core capabilities of the system: (i) answering physics questions using information contained within a single scientific publication, and (ii) performing cross-paper reasoning by combining evidence from multiple experimental analyses.
The experiments were conducted using three recent CMS analyses investigating different signatures of physics beyond the Standard Model (BSM):
\begin{itemize}
\item A search for heavy long-lived charged particles using ionization energy loss measurements \cite{CMS_HSCP_2024}
\item A search for light long-lived particles decaying to displaced jets \cite{CMS_LLP_2024}
\item A search for top squarks in final states with multiple jets and leptons \cite{CMS_STOP_2025}
\end{itemize}
These analyses were selected because they represent distinct classes of experimental signatures relevant to BSM physics, including detector-stable particles, displaced decays, and promptly decaying supersymmetric particles. Each analysis also contains heterogeneous scientific information such as numerical tables, experimental plots, and exclusion limits derived from collider data.
The associated HEPData entries accompanying these publications were ingested into the system's knowledge base, enabling retrieval of the numerical measurements underlying the experimental plots reported in the original analyses.

\subsection{Evaluation Methodology}

Evaluating systems designed for scientific literature analysis presents challenges that differ from traditional benchmark-based question answering tasks. Many of the queries considered in this study require interpretation of experimental measurements, reconstruction of physics plots from numerical datasets, or synthesis of information across multiple publications. Because these tasks produce structured explanations and visual outputs, conventional automated metrics such as exact-match accuracy are insufficient for assessing the quality of the generated results.

Instead, we adopt an \textit{LLM-as-a-Judge} evaluation strategy \cite{llmjudge}. In this framework, the responses generated by HEP-CoPilot are evaluated by multiple independent large language models acting as automated reviewers.

Specifically, we employ three different state-of-the-art LLMs—ChatGPT (OpenAI), Gemini (Google), and Claude (Anthropic)—to assess the quality of the generated outputs. Each model independently evaluates the responses according to predefined criteria including factual correctness, consistency with retrieved experimental evidence, clarity of explanation, and scientific coherence. Using multiple independent judges helps mitigate potential evaluation bias associated with single-model assessment.

To further contextualize the evaluation, we compare the responses generated by HEP-CoPilot with answers produced by directly querying these models using the original research papers in PDF format. This comparison provides a baseline representing standard document-based prompting without the proposed retrieval-augmented multi-agent reasoning pipeline.

Across the evaluated tasks, the responses generated by HEP-CoPilot consistently received higher evaluation scores from the independent LLM judges, indicating improved grounding in experimental data and more accurate interpretation of collider physics results relative to direct PDF-based querying.

\subsection{Single-Paper Physics Question Answering}

The first set of experiments evaluates the system's ability to retrieve experimental measurements, reconstruct physics plots, and generate evidence-grounded explanations using information contained within a single scientific publication. To improve the clarity of the evaluation, tasks are grouped according to the experimental analysis from which the information is retrieved.

\subsubsection{HSCP Search Analysis}

The first group of tasks is based on the heavy stable charged particle (HSCP) search \cite{CMS_HSCP_2024}.

\textbf{Q1.} Using the HEPData numerical points, plot the observed 95\% CL upper limits on the production cross section as a function of mass for gluino R-hadron, stop R-hadron, and stau models. Identify which model provides the strongest (lowest) limit.

\begin{figure}[!t]
\centering
\begin{subfigure}{0.49\textwidth}
\centering
\includegraphics[width=\linewidth]{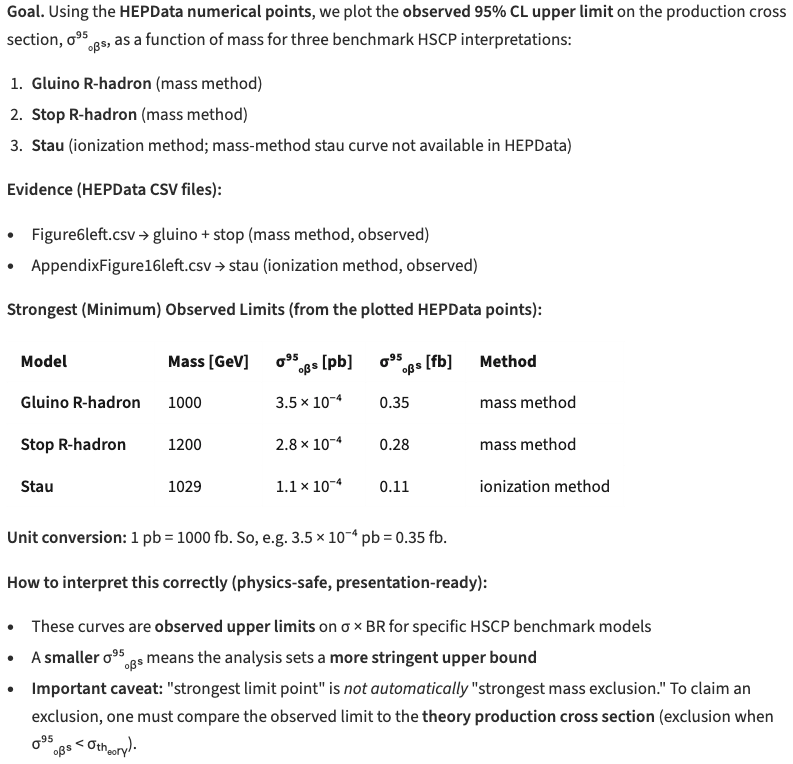}
\caption{Generated answer.}
\end{subfigure}
\hfill
\begin{subfigure}{0.49\textwidth}
\centering
\includegraphics[width=\linewidth]{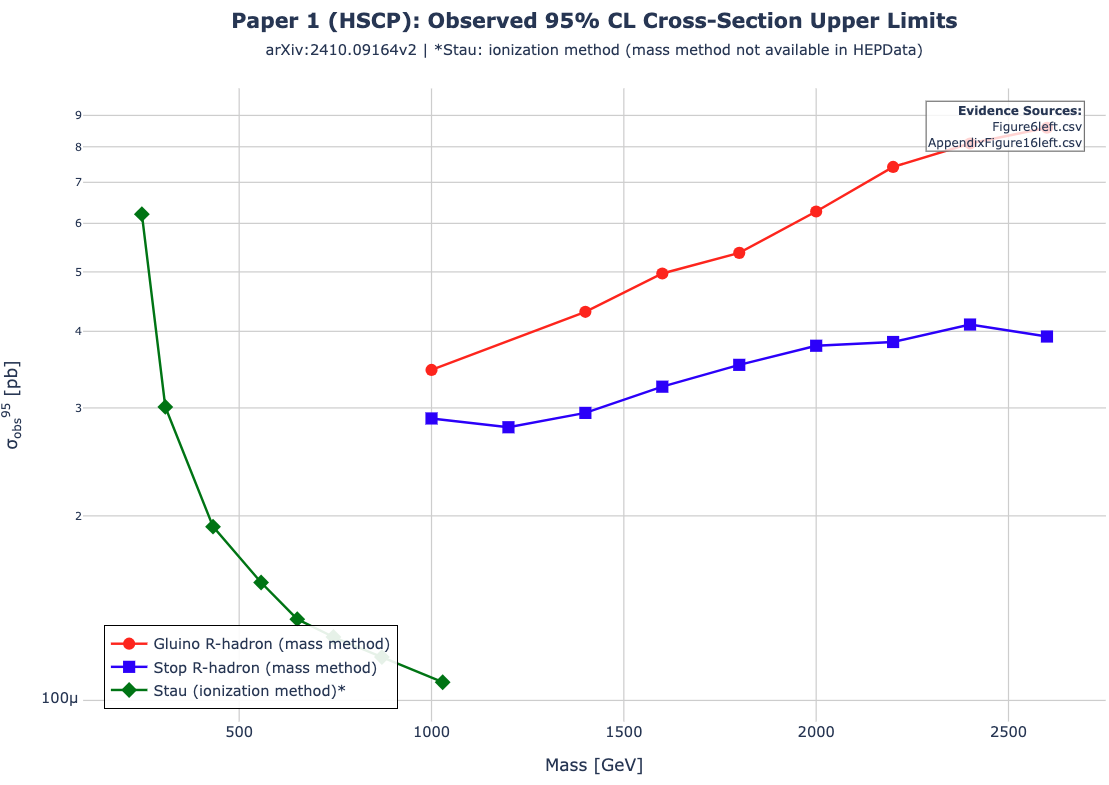}
\caption{Observed 95\% CL upper limits on the HSCP production cross section reconstructed from HEPData numerical records for gluino R-hadron, stop R-hadron, and stau benchmark models. The curves reproduce the exclusion-limit trends reported in the CMS HSCP analysis.}
\end{subfigure}
\caption{Result of Q1 illustrating reconstruction of HSCP cross-section limits.}
\label{fig:eval_single_1}
\end{figure}

\noindent\textbf{Interpretation (Fig.~\ref{fig:eval_single_1}).}  
The system reconstructs the observed cross-section limit curves directly from the HEPData numerical records and correctly identifies the model yielding the most stringent constraint. 

The reconstructed curves reproduce the qualitative behavior reported in the CMS analysis, where the stau interpretation provides the lowest cross-section upper limits within the accessible mass range.

This example illustrates that the framework can retrieve numerical measurements from HEPData and use them to perform physics-aware comparisons between different benchmark HSCP scenarios.

\bigskip

\textbf{Q2.} Explain the physical meaning of the pixel-only ionization discriminant requirement $F^{i}_{\mathrm{Pixels}} > 0.3$. Why does this selection suppress background in HSCP searches, and what types of tracks or detector effects does it reject?

\begin{figure}[!t]
\centering
\includegraphics[width=\linewidth]{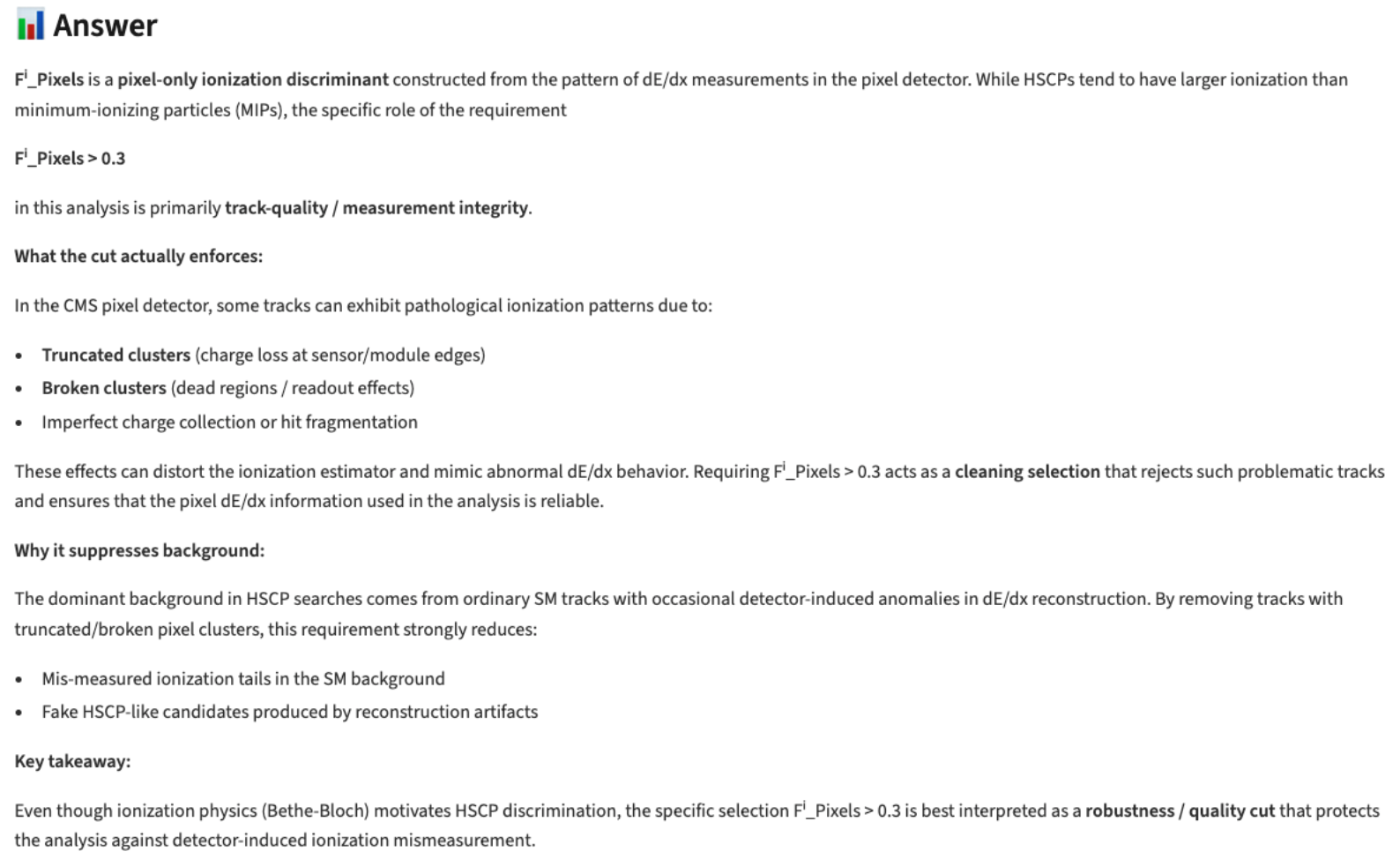}
\caption{Result of Q2 illustrating interpretation of the pixel-only ionization discriminant used in the HSCP search. 
The system explains the physical role of the $F^{i}_{\mathrm{Pixels}}$ requirement as a track-quality selection ensuring reliable pixel $dE/dx$ measurements and suppressing detector-induced ionization artifacts.}
\label{fig:eval_single_4}
\end{figure}

\noindent\textbf{Interpretation (Fig.~\ref{fig:eval_single_4}).}  
The system correctly identifies the requirement $F^{i}_{\mathrm{Pixels}} > 0.3$ as a selection related to the consistency of pixel-based ionization measurements used in the HSCP analysis.

While HSCP candidates are expected to exhibit anomalously large ionization energy loss ($dE/dx$) compared to minimum-ionizing particles, detector effects such as truncated pixel clusters, dead sensor regions, or imperfect charge collection can produce spurious high-ionization signals.

The $F^{i}_{\mathrm{Pixels}}$ discriminator evaluates the compatibility of the measured pixel $dE/dx$ pattern with a genuine high-ionization track. Requiring $F^{i}_{\mathrm{Pixels}} > 0.3$ therefore acts as a quality and robustness selection that suppresses tracks affected by detector artifacts while preserving genuine HSCP-like ionization signatures.

\subsubsection{Displaced-Jet LLP Search}

The second group of tasks is based on the displaced-jet long-lived particle search \cite{CMS_LLP_2024}.

\textbf{Q3.} Using the HEPData numerical measurements, plot the 95\% CL lower limit on the scalar mass $m_S$ as a function of the proper decay length $c\tau_0$ for the decay channels $S \rightarrow b\bar{b}$ and $S \rightarrow d\bar{d}$. Identify the lifetime region where the search sensitivity is weakest.

\begin{figure}[!t]
\centering
\begin{subfigure}{0.49\textwidth}
\centering
\includegraphics[width=\linewidth]{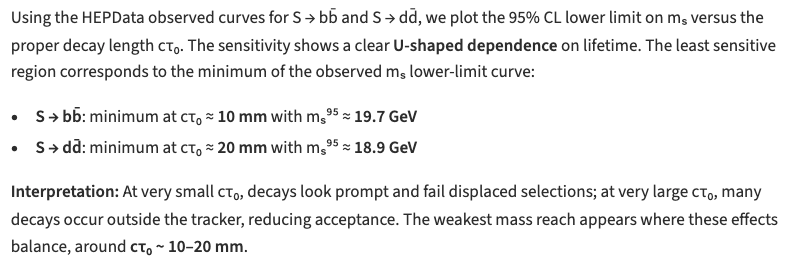}
\caption{Generated answer.}
\end{subfigure}
\hfill
\begin{subfigure}{0.49\textwidth}
\centering
\includegraphics[width=\linewidth]{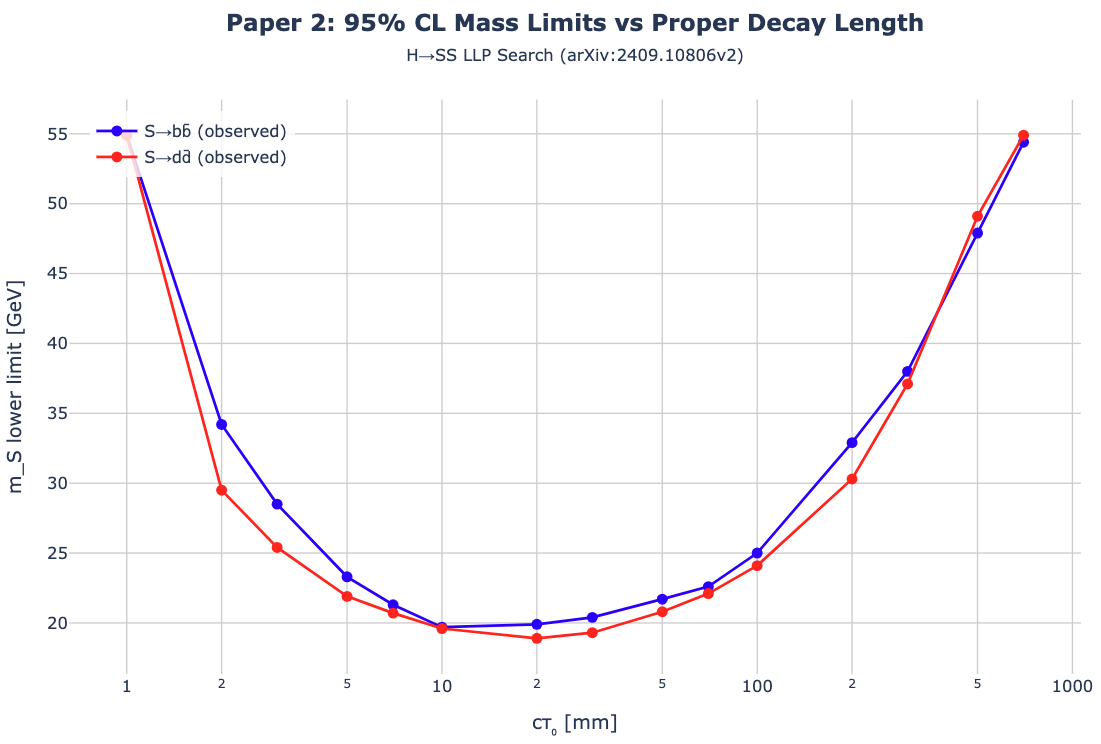}
\caption{Reconstructed $m_S$ limits vs $c\tau_0$.}
\end{subfigure}
\caption{Result of Q3 illustrating lifetime-dependent LLP mass limits.}
\label{fig:eval_single_2}
\end{figure}

\noindent\textbf{Interpretation (Fig.~\ref{fig:eval_single_2}).}  
The reconstructed curves reveal a characteristic lifetime dependence of the experimental sensitivity. The weakest limits occur in the intermediate lifetime region, where the decay vertices are displaced but still located within regions of the detector where reconstruction efficiency is reduced.

For very short lifetimes, the decays appear effectively prompt and therefore fail the displaced-vertex selection requirements. For very long lifetimes, a significant fraction of particles decay outside the tracker volume, again reducing the probability of reconstructing a displaced vertex.

The minimum mass reach therefore occurs at intermediate lifetimes ($c\tau_0 \approx 10$--$20\,\mathrm{mm}$), where these competing effects balance.

\bigskip

\textbf{Q4.} Using the HEPData points corresponding to Fig.~3 of the analysis, plot the observed data and the predicted background as a function of $g_{\mathrm{displaced}}$. For the highest $g_{\mathrm{displaced}}$ bin (signal-region-like bin), compute and report the numerical difference between the observed and predicted event yields.

\begin{figure}[!t]
\centering
\begin{subfigure}{0.49\textwidth}
\centering
\includegraphics[width=\linewidth]{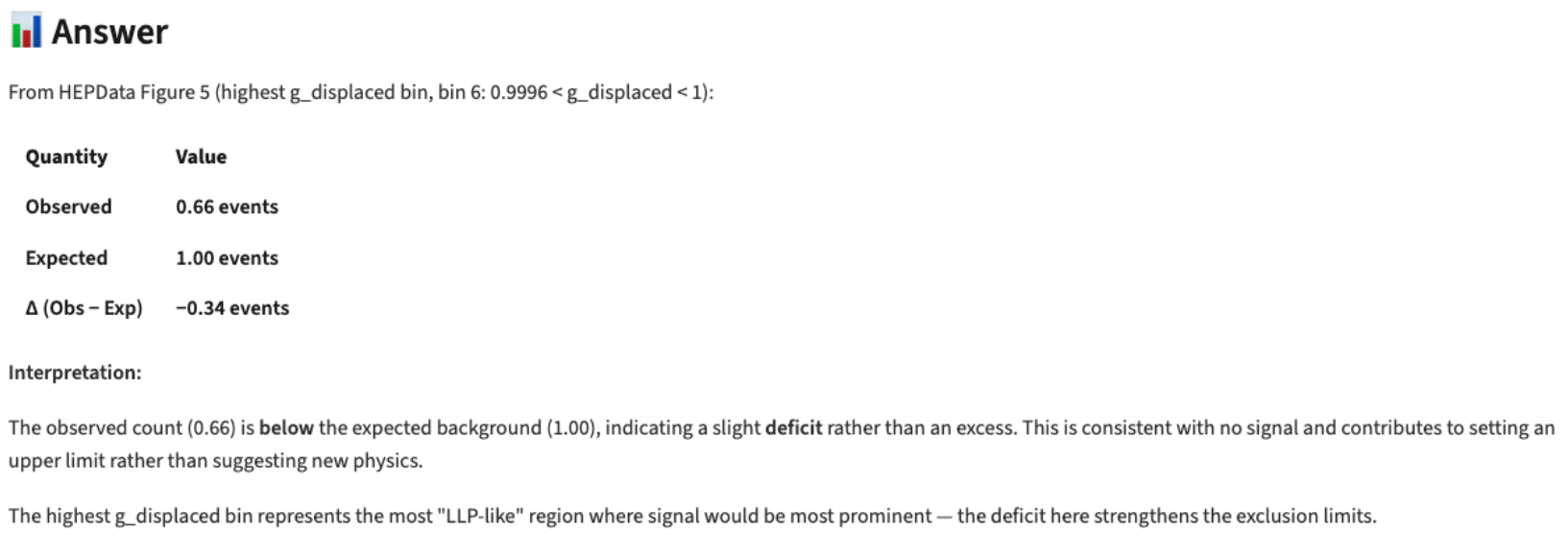}
\caption{Generated numerical summary and interpretation.}
\end{subfigure}
\hfill
\begin{subfigure}{0.49\textwidth}
\centering
\includegraphics[width=\linewidth]{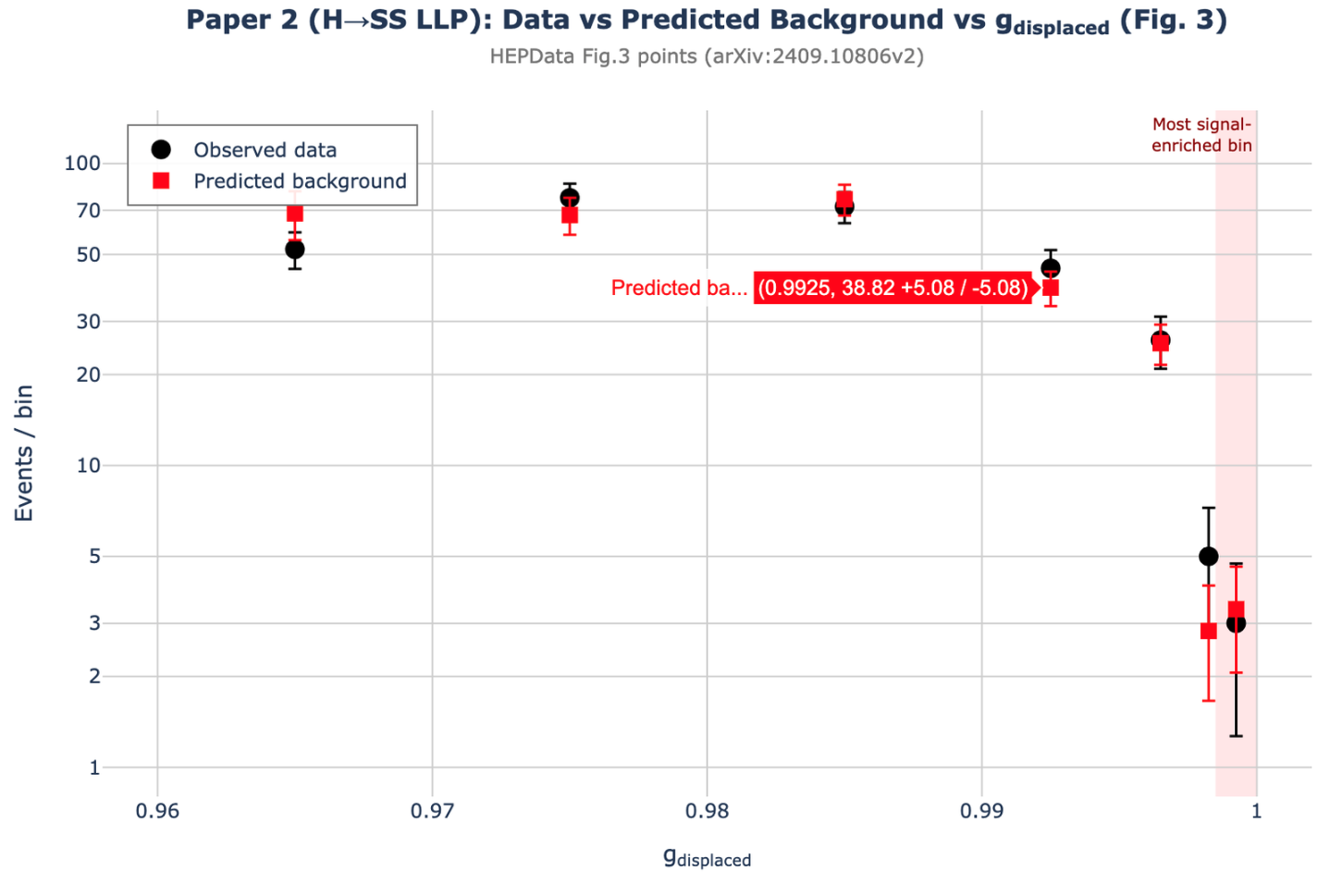}
\caption{Reconstructed observed and predicted yields vs $g_{\mathrm{displaced}}$.}
\end{subfigure}
\caption{Result of Q4 illustrating reconstruction of the $g_{\mathrm{displaced}}$ distribution and comparison between observed data and predicted background in the LLP search.}
\label{fig:eval_single_5}
\end{figure}

\noindent\textbf{Interpretation (Fig.~\ref{fig:eval_single_5}).}  
Using numerical measurements retrieved from HEPData, the system reconstructs the distribution of observed events and predicted background as a function of the displacement discriminant $g_{\mathrm{displaced}}$.

In the highest $g_{\mathrm{displaced}}$ bin, corresponding to the most signal-like region of the analysis, the observed yield is approximately $0.66$ events compared to an expected background of about $1.00$ events, giving $\Delta(\mathrm{Obs} - \mathrm{Exp}) \approx -0.34$.

The observed count is therefore slightly below the background prediction and does not indicate a statistically significant excess. This behavior is consistent with the background-only hypothesis and reflects the typical statistical fluctuations expected in low-yield signal regions of collider searches.

This example demonstrates that the framework can reconstruct signal-region distributions from HEPData and perform physics-aware interpretation of observed and predicted event yields.

\subsubsection{Top Squark Search}

The final group of single-paper tasks is based on the stop search analysis \cite{CMS_STOP_2025}.

\textbf{Q5.} Compare the observed cross-section limits as a function of the stop mass $m_{\tilde{t}}$ between the SYY $1\ell$ and RPV $1\ell$ channels. Which channel provides stronger limits?

\begin{figure}[!t]
\centering
\begin{subfigure}{0.49\textwidth}
\centering
\includegraphics[width=\linewidth]{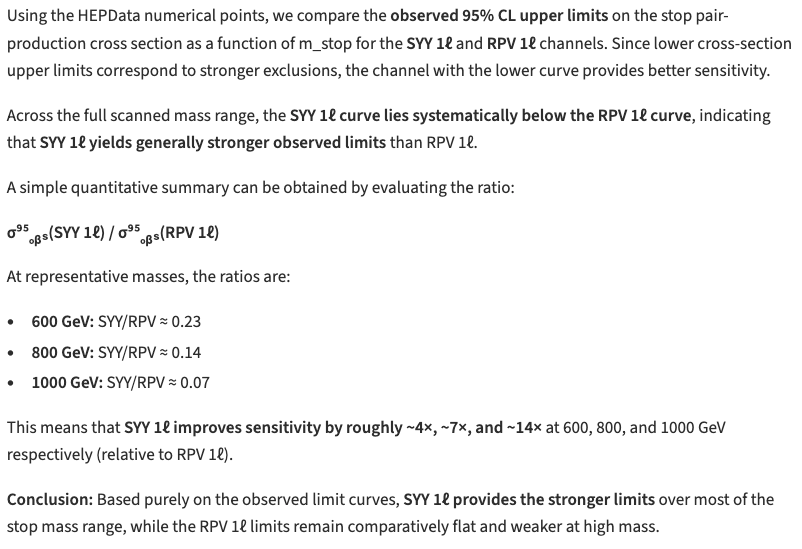}
\caption{Generated comparison answer.}
\end{subfigure}
\hfill
\begin{subfigure}{0.49\textwidth}
\centering
\includegraphics[width=\linewidth]{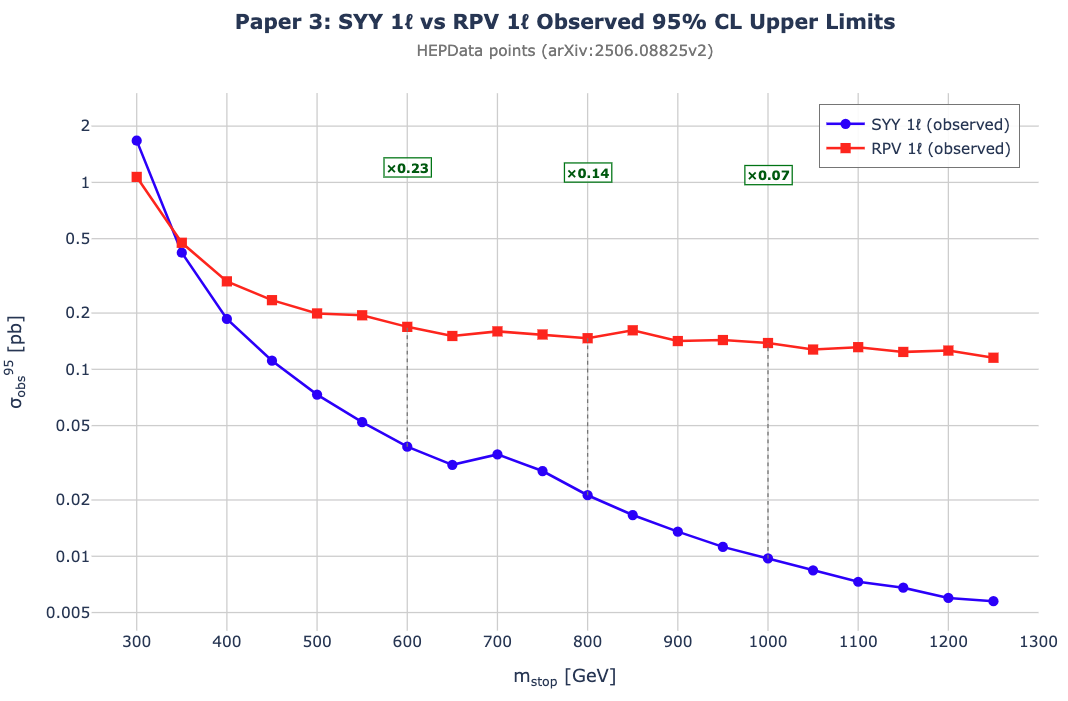}
\caption{Overlay of observed $\sigma^{95}$ limit curves.}
\end{subfigure}
\caption{Result of Q5 illustrating channel-to-channel comparison of stop search limits.}
\label{fig:eval_single_3}
\end{figure}

\noindent\textbf{Interpretation (Fig.~\ref{fig:eval_single_3}).}  
The system retrieves the observed cross-section limit curves for both channels and performs a direct comparison across the scanned stop-mass range. Since stronger limits correspond to lower cross-section upper bounds, the comparison can be performed directly at the level of the observed $\sigma^{95}$ curves.

Across the considered mass range, the SYY $1\ell$ limits lie systematically below those obtained in the RPV $1\ell$ channel, indicating stronger sensitivity in the SYY topology.

Quantitatively, the ratio $\sigma^{95}_{\mathrm{obs}}(\mathrm{SYY}) / \sigma^{95}_{\mathrm{obs}}(\mathrm{RPV})$ decreases with increasing stop mass, showing that the improvement in sensitivity becomes more pronounced at higher masses.

This comparison illustrates how the framework can retrieve numerical measurements from HEPData and perform physics-aware comparisons between different analysis channels within the same search.

\bigskip

\textbf{Q6.} Using the HEPData numerical points, plot the observed 95\% CL upper limit on the stop production cross section as a function of $m_{\tilde{t}}$ for the RPV $2\ell$ channel. Then extract the observed limit value at $m_{\tilde{t}} = 800$ GeV.

\begin{figure}[!t]
\centering
\begin{subfigure}{0.49\textwidth}
\centering
\includegraphics[width=\linewidth]{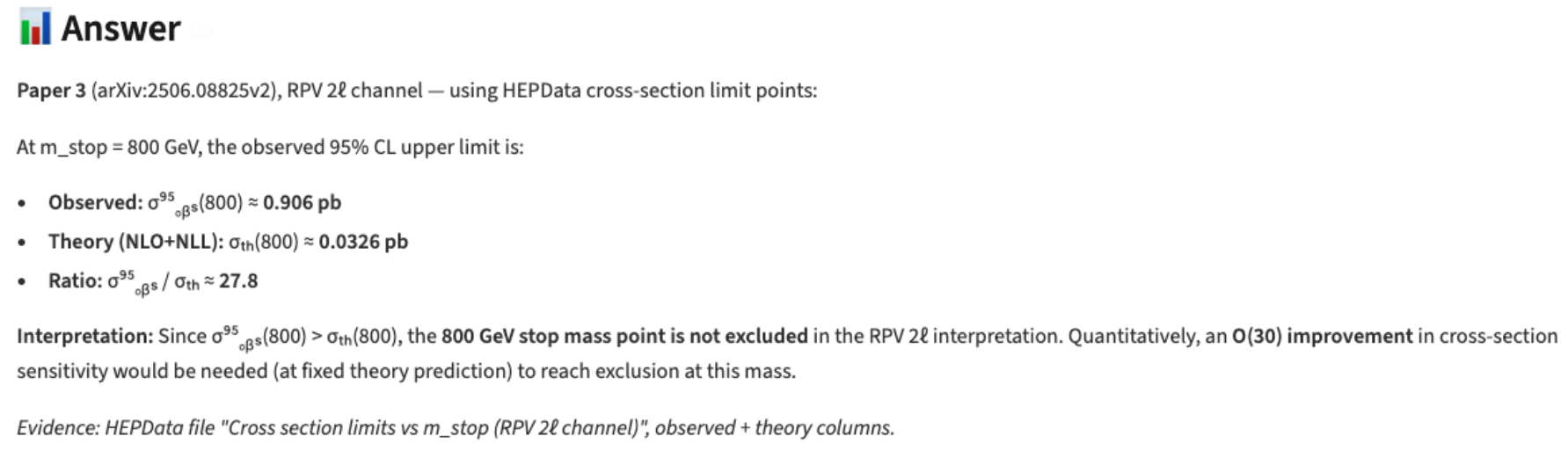}
\caption{Generated numerical extraction and interpretation.}
\end{subfigure}
\hfill
\begin{subfigure}{0.49\textwidth}
\centering
\includegraphics[width=\linewidth]{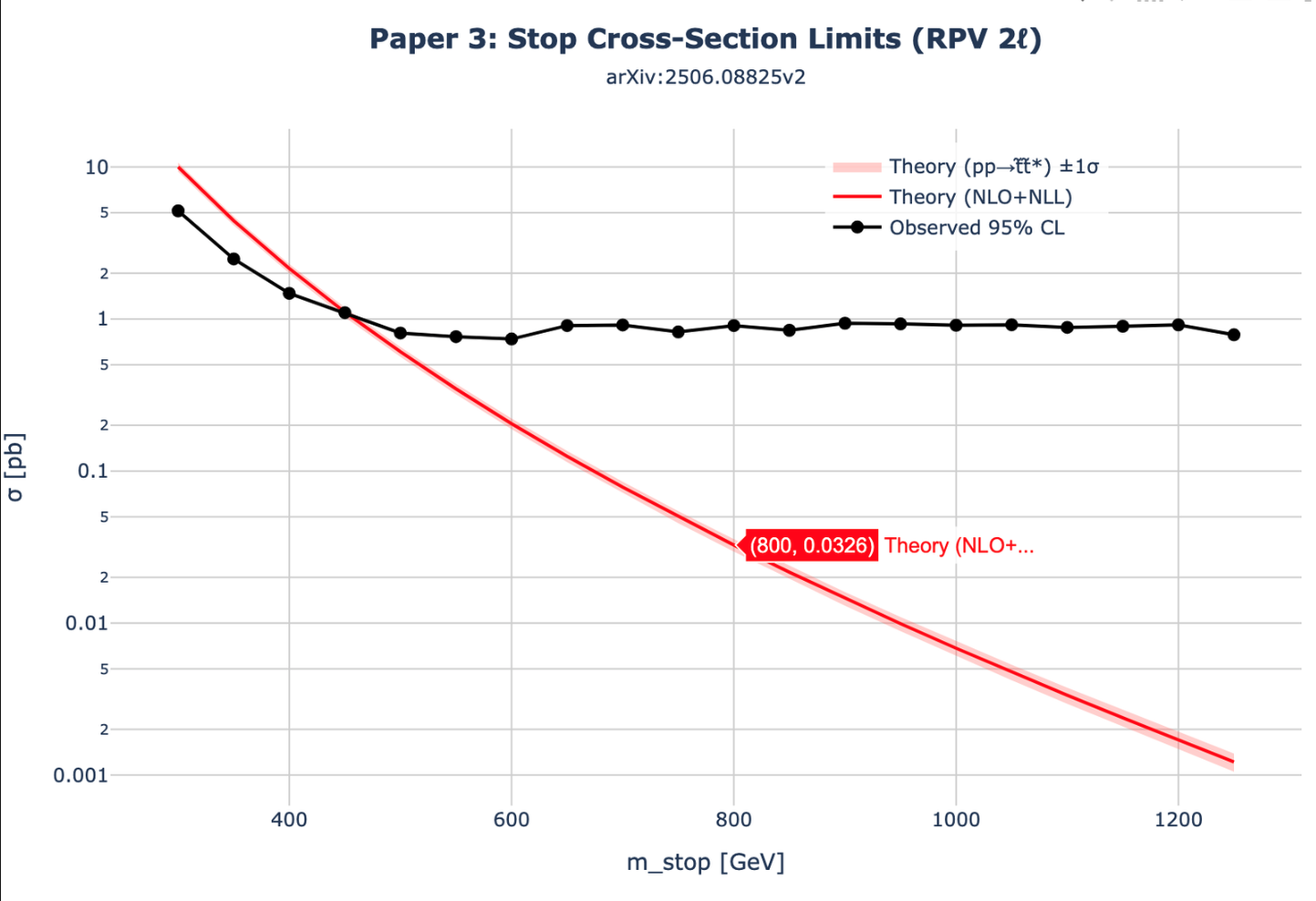}
\caption{Observed 95\% CL upper limits on the stop pair-production cross section as a function of the stop mass $m_{\tilde t}$ for the RPV $2\ell$ channel, reconstructed from HEPData numerical measurements. The theoretical production cross section ($pp \to \tilde t \tilde t^*$) calculated at NLO+NLL accuracy is shown for comparison.}
\end{subfigure}
\caption{Result of Q6 illustrating reconstruction of the stop production cross-section limits and extraction of the observed limit at $m_{\tilde{t}} = 800$ GeV for the RPV $2\ell$ channel.}
\label{fig:eval_single_6}
\end{figure}

\noindent\textbf{Interpretation (Fig.~\ref{fig:eval_single_6}).}  
The system reconstructs the observed cross-section limit curve for the RPV $2\ell$ stop search using numerical measurements retrieved from HEPData. From this reconstructed dataset, the observed 95\% CL upper limit at $m_{\tilde t} = 800$ GeV is extracted as approximately $\sigma^{95}_{\mathrm{obs}} \approx 0.906$ pb.

Comparing this value with the theoretical production cross section at the same mass point ($\sigma_{\mathrm{th}} \approx 0.0326$ pb) shows that the experimental upper limit remains significantly above the predicted signal rate. Consequently, the 800 GeV stop mass point is not excluded by this analysis.

Quantitatively, the ratio $\sigma^{95}_{\mathrm{obs}} / \sigma_{\mathrm{th}} \approx 28$ indicates that an improvement of roughly a factor of 30 in experimental sensitivity would be required to reach exclusion at this mass point.

\subsection{Cross-Paper Scientific Reasoning}

Beyond analysis of individual publications, HEP-CoPilot is designed to support reasoning across multiple experimental studies.

\textbf{Q7.} Overlay the observed 95\% CL cross-section limits from the HSCP gluino search and the stop RPV $2\ell$ channel on the same plot and compare their relative scales.

\begin{figure}[!t]
\centering
\begin{subfigure}{0.49\textwidth}
\centering
\includegraphics[width=\linewidth]{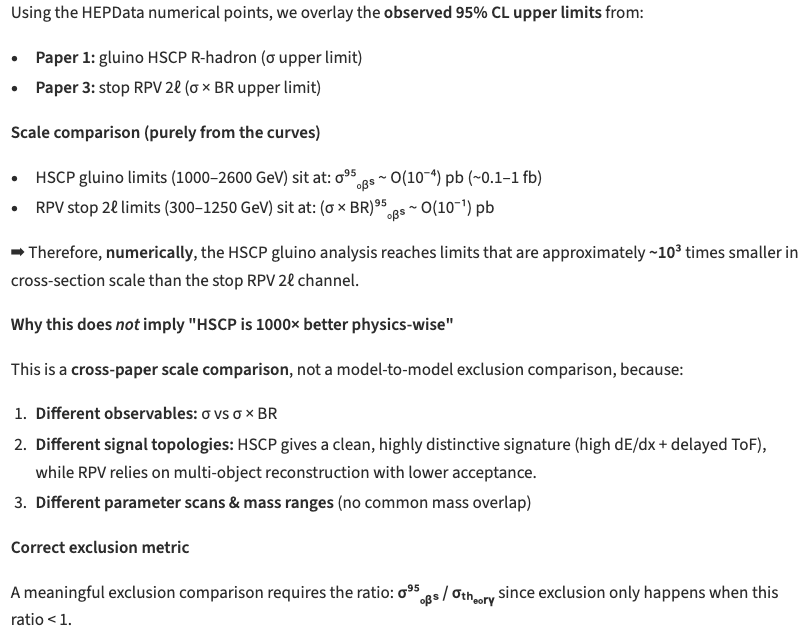}
\caption{Generated cross-paper explanation.}
\end{subfigure}
\hfill
\begin{subfigure}{0.49\textwidth}
\centering
\includegraphics[width=\linewidth]{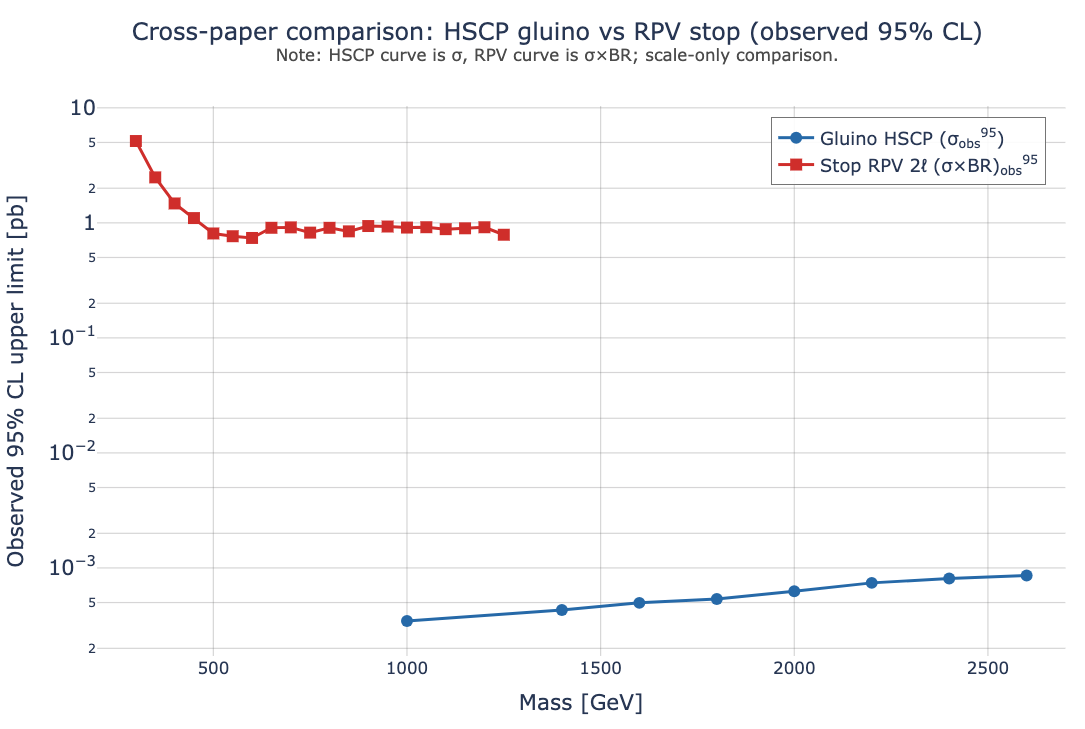}
\caption{Overlay of limits from two analyses.}
\end{subfigure}
\caption{Result of Q7 illustrating cross-paper comparison of observed limits.}
\label{fig:eval_cross_1}
\end{figure}

\noindent\textbf{Interpretation (Fig.~\ref{fig:eval_cross_1}).}  
The system retrieves numerical limit curves from both analyses and produces a combined visualization that allows direct comparison of their typical scales. The overlay shows that the HSCP gluino limits lie roughly three orders of magnitude below those obtained in the RPV stop analysis.

The generated explanation correctly emphasizes that such cross-paper comparisons must be interpreted with caution. The two analyses constrain different observables ($\sigma$ versus $\sigma \times \mathrm{BR}$), probe different signal topologies, and scan different mass ranges. Consequently, the comparison should be understood as a scale comparison of the reported limits rather than a direct comparison of experimental sensitivity or model exclusion power.

This example illustrates the framework’s ability to combine numerical information from multiple publications while preserving the physical context of the original analyses.

\bigskip

\textbf{Q8.} Classify the analyzed searches according to the lifetime regime of the targeted particles (prompt, displaced, detector-stable) and their dominant experimental signatures.

\begin{figure}[!t]
\centering
\begin{subfigure}{0.49\textwidth}
\centering
\includegraphics[width=\linewidth]{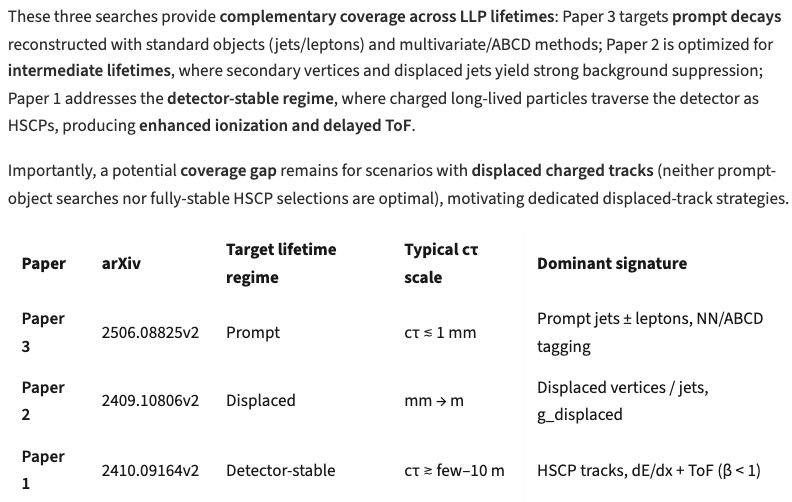}
\caption{Generated cross-paper explanation.}
\end{subfigure}
\hfill
\begin{subfigure}{0.49\textwidth}
\centering
\includegraphics[width=\linewidth]{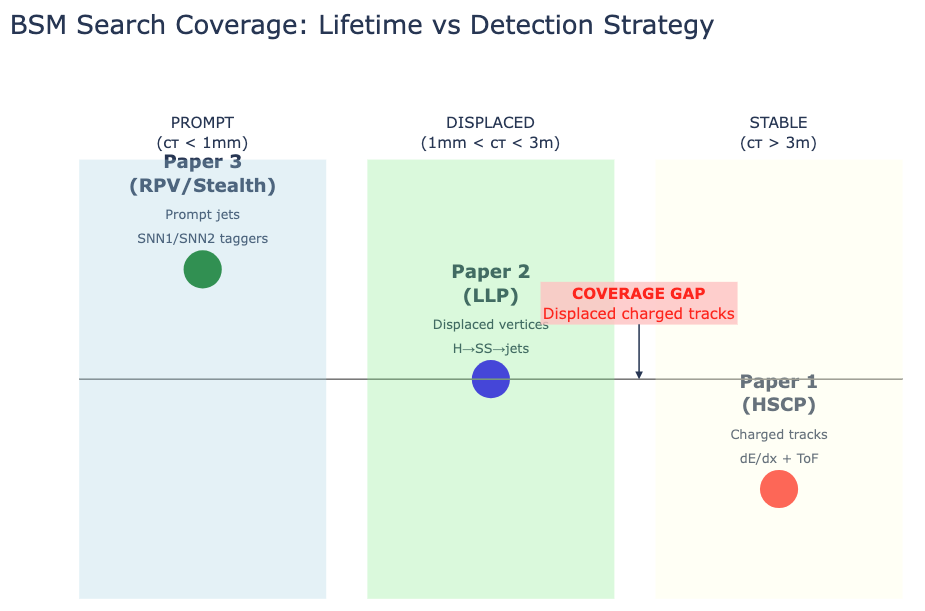}
\caption{Structured classification summary.}
\end{subfigure}
\caption{Result of Q8 illustrating classification of searches by lifetime regime and detector signature.}
\label{fig:eval_cross_2}
\end{figure}

\noindent\textbf{Interpretation (Fig.~\ref{fig:eval_cross_2}).}  
The system synthesizes information from multiple publications and organizes the searches according to the lifetime regime of the targeted particles and their dominant experimental signatures.

The resulting classification highlights the complementary coverage of prompt, displaced, and detector-stable searches. Prompt-object analyses are sensitive to very short lifetimes where decays occur close to the interaction point, while displaced-vertex searches target intermediate lifetimes where secondary vertices can be reconstructed within the tracker volume. HSCP analyses probe the opposite extreme in which long-lived charged particles traverse the detector without decaying.

This classification illustrates how different experimental strategies collectively cover a broad range of particle lifetimes in searches for physics beyond the Standard Model.

\bigskip

\textbf{Q9.} Paper 2 provides limits as a function of lifetime $c\tau_0$, while Paper 1 probes the effectively stable (“detector-stable”) regime. Using the Paper 2 HEPData curves, identify the lifetime region where sensitivity is weakest and discuss how this lifetime window complements or overlaps with the stable regime covered by Paper 1.

\begin{figure}[!t]
\centering
\begin{subfigure}{0.45\textwidth}
\centering
\includegraphics[width=\linewidth]{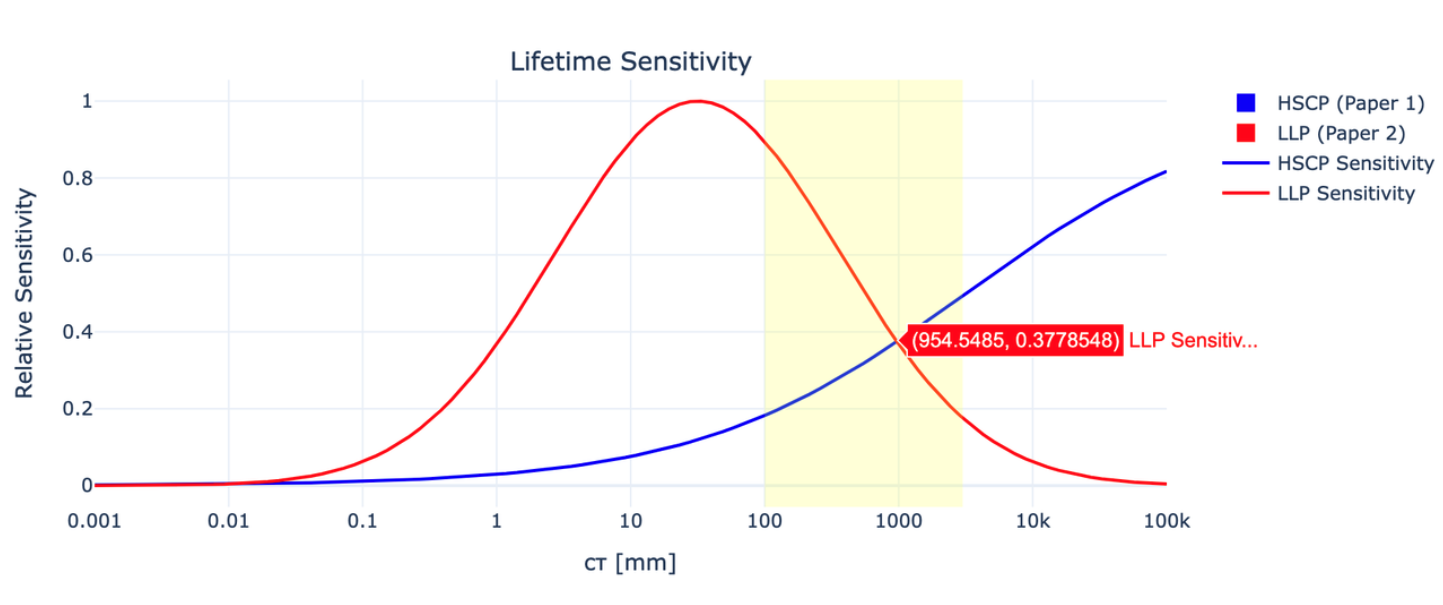}
\caption{Comparison of detector signatures.}
\end{subfigure}
\hfill
\begin{subfigure}{0.45\textwidth}
\centering
\includegraphics[width=\linewidth]{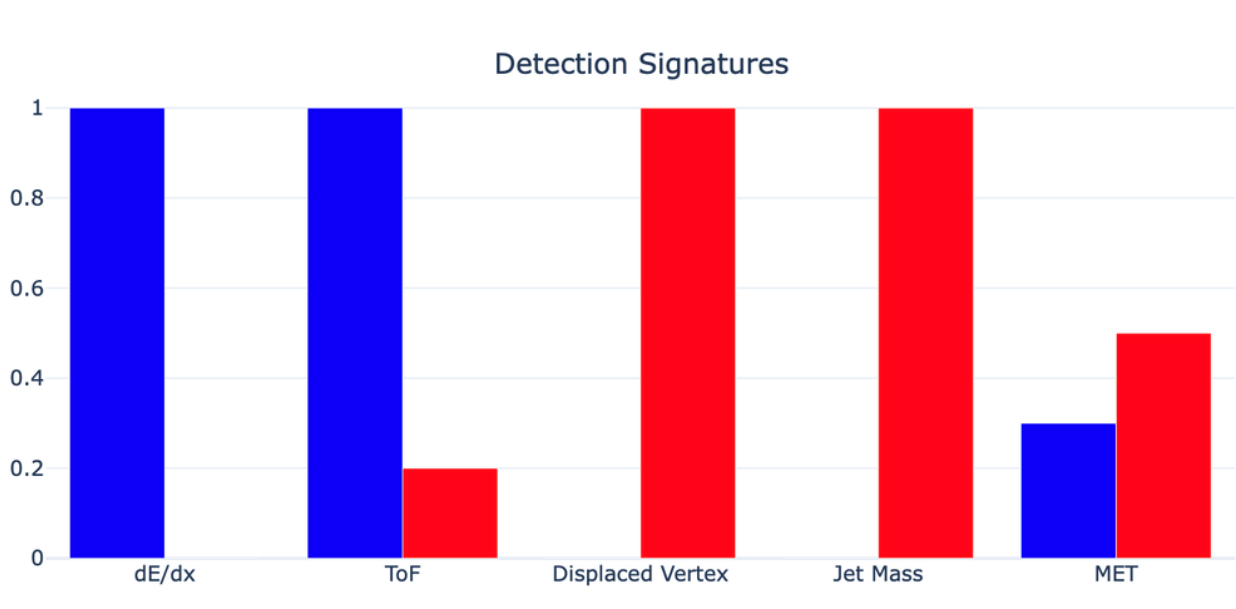}
\caption{Lifetime sensitivity comparison.}
\end{subfigure}
\hfill
\begin{subfigure}{0.45\textwidth}
\centering
\includegraphics[width=\linewidth]{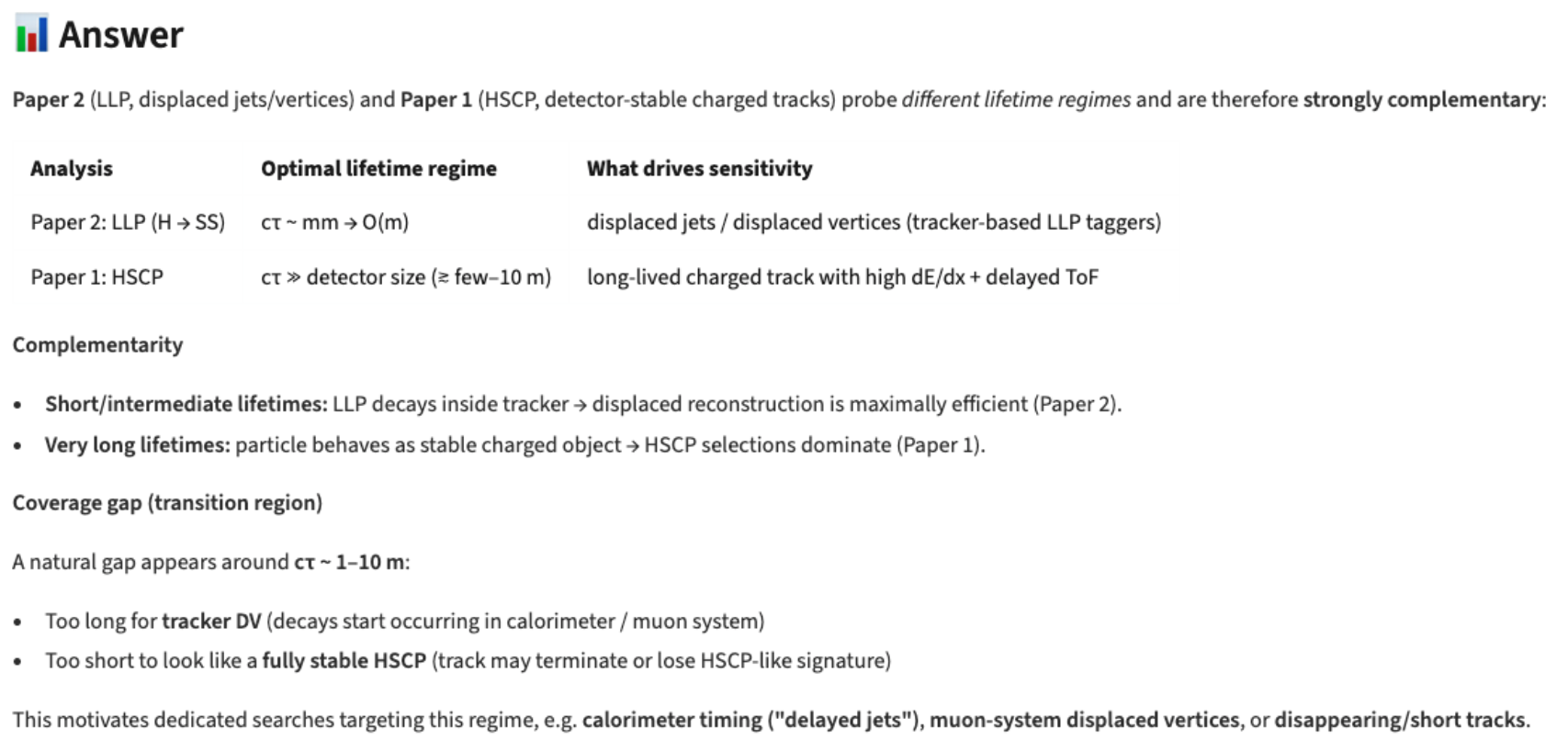}
\caption{Generated cross-paper interpretation.}
\end{subfigure}

\caption{Result of Q9 illustrating cross-paper reasoning about lifetime-dependent sensitivity in LLP and HSCP searches.}
\label{fig:eval_cross_3}
\end{figure}

\noindent\textbf{Interpretation (Fig.~\ref{fig:eval_cross_3}).}  
The system analyzes the lifetime-dependent limits reported in the displaced-jet LLP search and identifies the region where experimental sensitivity becomes weakest. In the displaced-vertex analysis, the sensitivity decreases when the particle lifetime becomes sufficiently large that decays begin to occur outside the tracker volume.

This transition region approximately corresponds to lifetimes of order $c\tau \sim 1$--$10\,\mathrm{m}$.
In this regime, the efficiency of displaced-vertex reconstruction in the LLP search decreases significantly, while the particle lifetime remains too short for the particle to appear as a detector-stable charged track in HSCP analyses.

Consequently, this region lies between the optimal sensitivity regimes of the displaced LLP search (Paper 2) and the HSCP search (Paper 1). The two analyses therefore provide complementary coverage of the lifetime parameter space, while leaving a partially uncovered transition region that motivates dedicated searches for displaced charged tracks.

\bigskip

\textbf{Q10.} Considering the three searches together (HSCP detector-stable charged particles, displaced-jet LLP search, and prompt stop search with ABCDisCoTEC), identify potential coverage gaps in particle lifetime and experimental signatures. Propose concrete search strategies that could improve sensitivity in the missing regions.

\begin{figure}[!t]
\centering
\begin{subfigure}{0.49\textwidth}
\centering
\includegraphics[width=\linewidth]{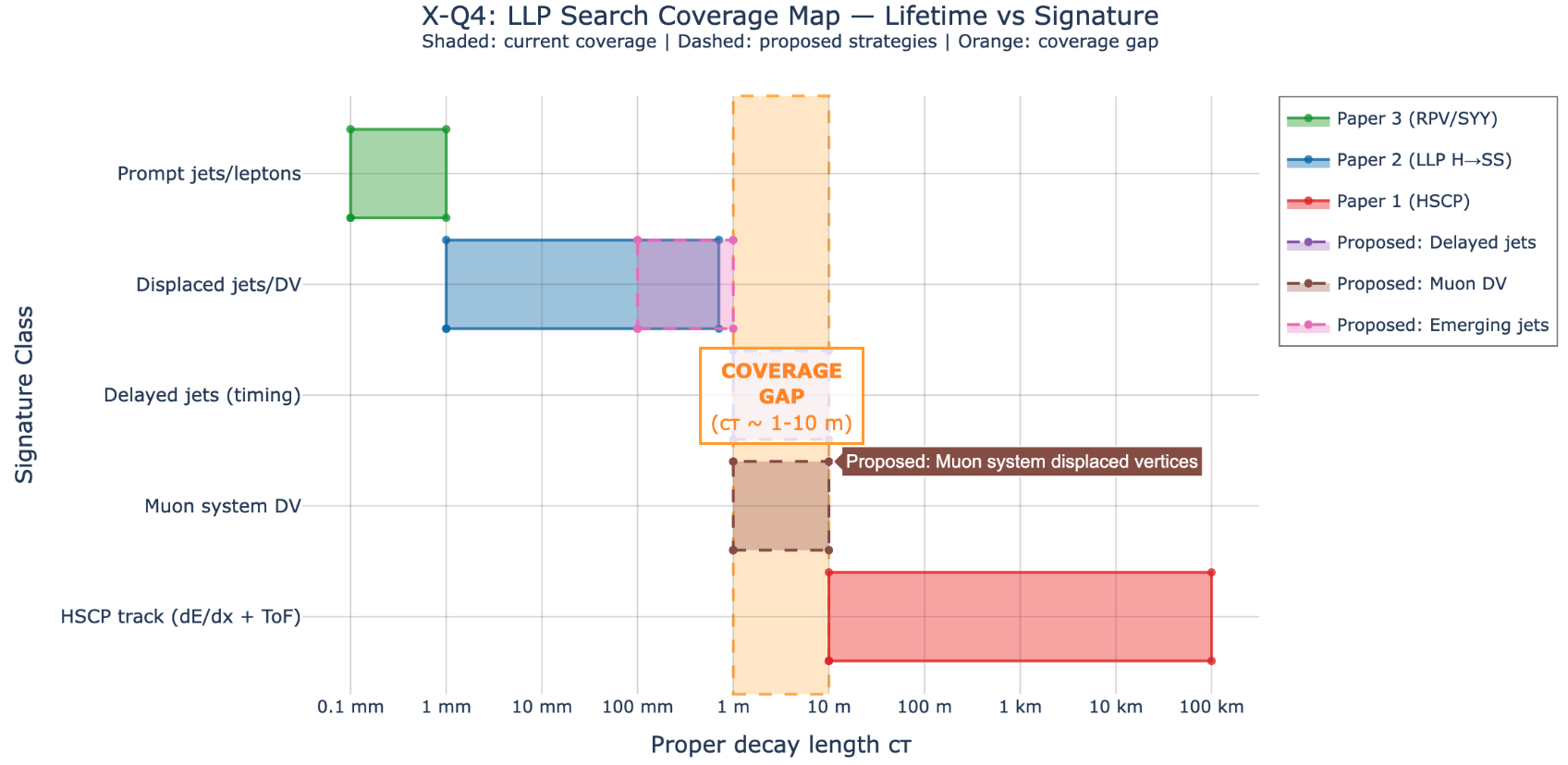}
\caption{Lifetime–signature coverage map derived from the three analyses.}
\end{subfigure}
\hfill
\begin{subfigure}{0.49\textwidth}
\centering
\includegraphics[width=\linewidth]{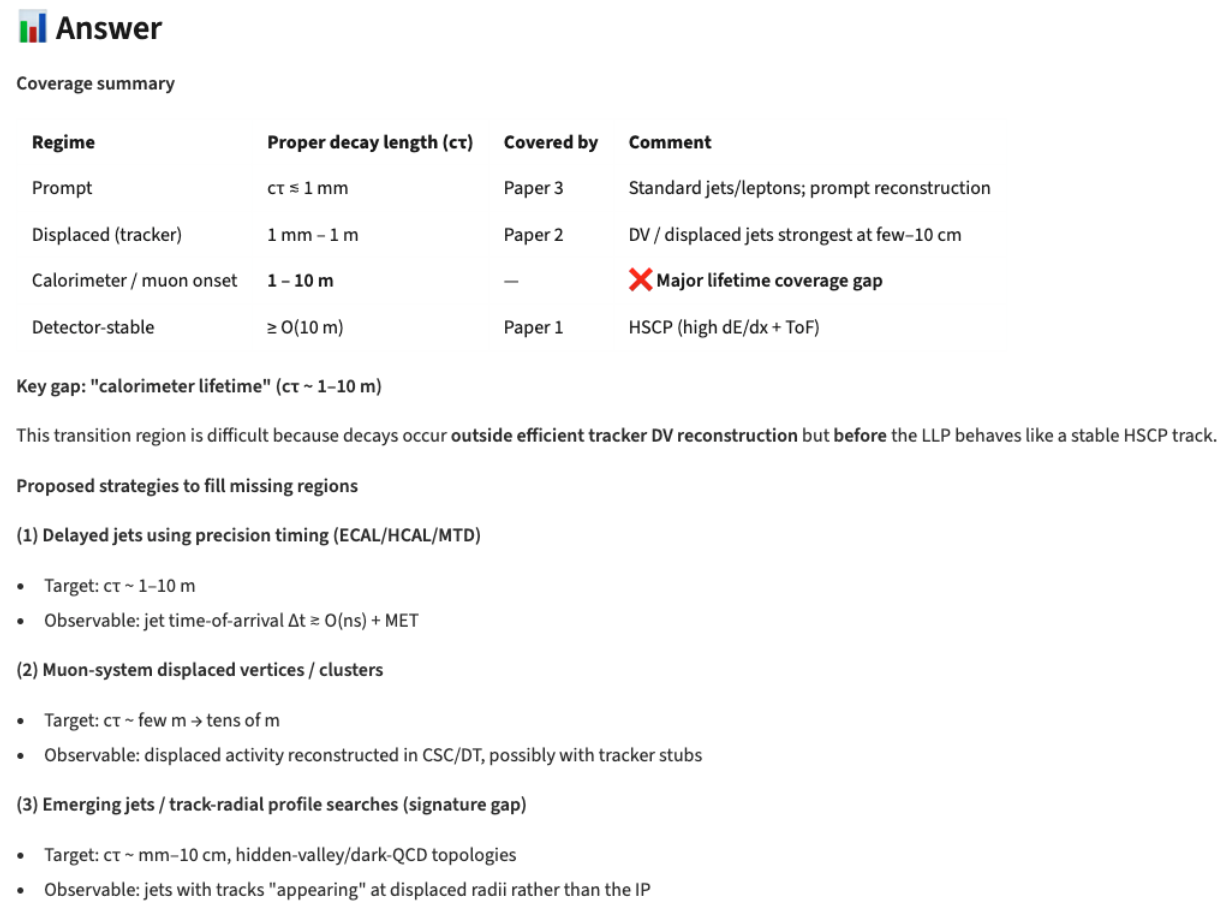}
\caption{Generated reasoning and proposed search strategies.}
\end{subfigure}

\caption{Lifetime–signature coverage map constructed by combining information from the three analyzed searches. The shaded regions represent the parameter space currently covered by the prompt, displaced-vertex, and HSCP analyses, while dashed regions illustrate proposed search strategies that could improve sensitivity in uncovered lifetime regimes.}
\label{fig:eval_cross_4}
\end{figure}

\noindent\textbf{Interpretation (Fig.~\ref{fig:eval_cross_4}).}  
By synthesizing information from the three analyses, the system constructs a qualitative coverage map of experimental sensitivity as a function of particle lifetime and detector signature. The resulting map highlights how prompt-object searches, displaced-vertex analyses, and HSCP searches probe complementary regions of the lifetime parameter space.

The visualization also reveals a transition region around $c\tau \sim 1$--$10\,\mathrm{m}$, where the efficiency of tracker-based displaced-vertex reconstruction decreases, while the particle lifetime remains too short for the particle to appear as a detector-stable HSCP track. This region therefore represents a potential coverage gap in current search strategies.

Based on this observation, the system proposes several experimental strategies that could improve sensitivity in this regime, including delayed-jet searches using precision timing information, displaced-vertex reconstruction in the muon system, and emerging-jet analyses targeting hidden-sector signatures. These proposals illustrate the framework's ability not only to aggregate information from multiple publications but also to generate physics-motivated hypotheses about unexplored regions of parameter space.

\subsection{Overall Performance}
To quantitatively assess the quality of the generated responses, we adopt an LLM-based evaluation strategy following the LLM-as-a-Judge methodology
In this framework, responses generated by HEP-CoPilot are evaluated by multiple independent language models that assess their correctness, completeness, and scientific consistency relative to the ground-truth information contained in the source publications.

Each response is scored according to predefined evaluation criteria including factual correctness, consistency with retrieved evidence, and clarity of the generated explanation. 
The evaluation results indicate that the system consistently produces scientifically grounded responses supported by experimental data retrieved from the literature.

Overall, the experimental results demonstrate that the proposed system is capable of retrieving experimental measurements, reconstructing physics plots, and synthesizing cross-paper insights. 
These capabilities highlight the potential of AI-assisted systems to support literature exploration and knowledge discovery in high-energy physics research.

\section{Discussion}

The rapid growth of experimental results produced by modern collider experiments has significantly increased the complexity of navigating high-energy physics literature. Contemporary searches for physics beyond the Standard Model (BSM) involve detailed analyses combining detector data, statistical interpretations, numerical measurements, and graphical representations of exclusion limits. As a consequence, extracting meaningful insights from the literature often requires researchers to manually integrate information distributed across multiple publications, tables, figures, and external data repositories.

While machine learning techniques have been widely adopted within particle physics for tasks such as event classification, detector reconstruction, and anomaly detection, comparatively little work has focused on assisting physicists in navigating and interpreting the scientific literature itself. Existing language-model-based systems typically operate as general-purpose text assistants and therefore lack the ability to reason over the heterogeneous information structures characteristic of particle physics publications.

The HEP-CoPilot framework introduced in this work addresses this gap by combining retrieval-augmented language models with a modular multi-agent reasoning architecture designed specifically for high-energy physics literature. By integrating textual information from scientific papers with structured numerical data from HEPData and reconstructed experimental plots, the system enables multimodal reasoning over the experimental results reported in collider analyses.

The experimental case studies demonstrate several capabilities that are particularly relevant for particle physics research workflows. First, the system is able to retrieve experimental measurements and reconstruct physics plots directly from numerical datasets. This capability allows the system not only to reproduce published plots but also to use the reconstructed numerical measurements for comparative reasoning across multiple collider analyses. An important capability of the framework is the ability to overlay and compare exclusion limits originating from different experimental analyses. By reconstructing plots directly from numerical datasets, the system can generate combined visualizations in which limits from multiple collider searches are displayed within a single coordinate system. This functionality enables rapid comparison of experimental sensitivities across different analyses and facilitates phenomenological reasoning and reinterpretation of collider search limits.

Unlike conventional literature retrieval systems that operate purely on textual information, HEP-CoPilot is capable of reconstructing experimental exclusion limits directly from numerical datasets and using these reconstructed measurements to support physics-aware reasoning across multiple publications.

An important aspect of the proposed framework is that it operates directly on the numerical measurements underlying experimental plots rather than relying solely on textual descriptions. In collider physics, exclusion limits are typically presented as graphical summaries of complex statistical analyses. By reconstructing these limits from structured HEPData records, the system gains access to the underlying numerical measurements, enabling quantitative comparison of experimental constraints and supporting physics-aware reasoning across multiple analyses.

These capabilities suggest that systems such as HEP-CoPilot may function as \emph{scientific co-pilots} for researchers working with complex experimental literature. Rather than replacing the role of physicists, the system is intended to augment the reasoning process by rapidly retrieving relevant experimental evidence, reconstructing numerical results, and organizing information across multiple analyses. In practice, such tools can support human–AI collaboration in which a physicist interacts with an AI assistant capable of rapidly exploring the literature, potentially accelerating the cycle of hypothesis evaluation, result interpretation, and comparative analysis across experiments.

From a broader perspective, the framework represents an initial step toward AI-assisted literature analysis in particle physics. As the number of experimental analyses and publicly available datasets continues to grow, tools capable of integrating heterogeneous sources of scientific information may become increasingly important for navigating the expanding body of collider physics results.

Despite these promising results, several limitations remain that provide directions for future research. First, the current study evaluates the system using a limited number of CMS analyses. This restriction is primarily due to computational constraints, as the prototype implementation was developed and tested on a local computing environment with limited storage and indexing capacity. Consequently, only three publications were incorporated into the experimental evaluation. Future work could extend the framework to much larger collections of particle physics publications by deploying the system on higher-capacity computing infrastructure capable of storing and indexing large-scale scientific corpora.

Second, the current implementation employs locally deployed language models through the Ollama framework. While this approach provides flexibility and enables offline experimentation, the reasoning capabilities of local models may be more limited compared to larger state-of-the-art systems such as ChatGPT or Claude. Integrating more advanced large language models could further improve the system's ability to generate detailed explanations and perform complex scientific reasoning over retrieved evidence.

Another limitation concerns the dependence on structured datasets such as those provided by HEPData. While many experimental analyses publish numerical results in machine-readable form, some information remains embedded in figures or complex tables within the papers themselves. Incorporating automated figure digitization and table extraction techniques could further improve the coverage of the system and allow it to operate on a broader range of publications.

In addition, the current reasoning framework focuses primarily on retrieving and synthesizing experimental information rather than performing deeper theoretical interpretation of experimental results. Extending the system with physics-aware reasoning modules capable of incorporating theoretical predictions, cross-section calculations, or global constraint comparisons could significantly expand its usefulness for phenomenological studies.

Finally, as with all language-model-based systems, ensuring reliability and minimizing the risk of hallucinated interpretations remains an important challenge. Although the retrieval-augmented design grounds responses in experimentally reported evidence, further work on verification mechanisms and uncertainty-aware reasoning may be necessary to ensure robust use of such systems in scientific workflows.

Despite these limitations, the results presented in this work demonstrate that combining retrieval-augmented reasoning with domain-aware multi-agent architectures can significantly improve the accessibility and interpretability of high-energy physics literature. By enabling structured exploration of experimental results and facilitating cross-paper comparisons, frameworks such as HEP-CoPilot may help researchers navigate increasingly complex bodies of scientific knowledge and accelerate the interpretation of new experimental findings.

Rather than replacing the role of physicists, systems such as HEP-CoPilot are designed to augment the scientific reasoning process. By rapidly retrieving experimental evidence, reconstructing numerical results, and organizing information across multiple publications, such systems may enable human researchers to explore the literature more efficiently and focus their efforts on higher-level scientific interpretation and hypothesis development. In this sense, HEP-CoPilot can be viewed as an early prototype of an AI-assisted scientific co-pilot capable of supporting physicists in navigating increasingly complex experimental landscapes.

\section{Conclusion}

In this work, we presented \textbf{HEP-CoPilot}, a multi-agent AI framework designed to support the exploration and interpretation of high-energy physics literature. The proposed system integrates retrieval-augmented language models with a modular reasoning architecture capable of processing heterogeneous scientific information, including textual descriptions, structured experimental datasets, and graphical representations of experimental limits.

Through a series of case studies based on recent CMS analyses, we demonstrated that the system can retrieve experimentally reported measurements, reconstruct physics plots directly from HEPData numerical records, and generate structured explanations grounded in retrieved scientific evidence. The framework further enables cross-paper reasoning by combining information from multiple analyses, allowing automated comparison of experimental constraints that are typically performed manually by researchers.

These results suggest that retrieval-augmented, domain-aware AI systems can provide an effective interface for navigating the increasingly complex landscape of particle physics literature. By assisting researchers in retrieving relevant experimental evidence, reconstructing numerical results, and synthesizing insights across multiple publications, systems such as HEP-CoPilot have the potential to function as scientific co-pilots that augment the reasoning process of physicists working with large bodies of experimental data.

As experimental results and scientific publications continue to grow in volume and complexity, tools that facilitate structured exploration of the literature may become increasingly important for accelerating the interpretation of new physics searches. The framework presented in this work represents an initial step toward AI-assisted literature analysis in high-energy physics and highlights the potential of multi-agent AI systems to support collaborative human–AI scientific workflows.

\section*{CRediT authorship contribution statement}

\textbf{Altan Cakir:} Supervision, Methodology, Writing – review \& editing.
\textbf{Ayca Yerlikaya:} Conceptualization, Methodology, Software, Writing.

\section*{Data availability}

The implementation of HEP-CoPilot is publicly available at:

\url{https://github.com/aycayk/HEP-CoPilot}


\begin{thebibliography}{99}

\bibitem{atlas2008}
ATLAS Collaboration,
The ATLAS Experiment at the CERN Large Hadron Collider,
Journal of Instrumentation 3 (2008) S08003.

\bibitem{cms2008}
CMS Collaboration,
The CMS Experiment at the CERN LHC,
Journal of Instrumentation 3 (2008) S08004.

\bibitem{denby1988}
B. Denby,
Neural Networks for Pattern Recognition in High-Energy Physics Events,
Computer Physics Communications 49 (1988) 429–448.

\bibitem{radovic2018ml}
A. Radovic et al.,
Machine Learning at the Energy and Intensity Frontiers of Particle Physics,
Nature 560 (2018) 41–48.

\bibitem{ramprasad2025}
R. Ramprasad et al.,
Large Language Models in Science,
arXiv:2501.05382 (2025).

\bibitem{lewis2020rag}
P. Lewis et al.,
Retrieval-Augmented Generation for Knowledge-Intensive NLP Tasks,
NeurIPS (2020), arXiv:2005.11401.

\bibitem{lala2023paperqa}
J. Lála, B. Albalawi, Z. Akata,
PaperQA: A Retrieval-Augmented Generative Agent for Scientific Research,
EMNLP (2023), arXiv:2307.07782.

\bibitem{pramanick2024spiqa}
S. Pramanick, S. Ghosh et al.,
SPIQA: Benchmarking Scientific Paper Information Extraction and Question Answering,
ACL Findings (2024), arXiv:2408.06292.

\bibitem{ghosh2023scitabqa}
S. Ghosh, S. Pramanick et al.,
SciTabQA: Tabular Reasoning in Scientific Papers,
EMNLP Findings (2023), arXiv:2305.07966.

\bibitem{simons2024astrohepbert}
B. Simons, N. P. Hartland,
Astro-HEP-BERT: Domain-Adaptive Language Models for Astronomy and High-Energy Physics,
arXiv:2401.02755 (2024).

\bibitem{hellert2024physbert}
T. Hellert et al.,
PhysBERT: A Text Embedding Model for Physics Literature,
arXiv:2403.08367 (2024).

\bibitem{nguyen2023astrollama}
H. Q. Nguyen et al.,
AstroLLaMA: An Adaptive Large Language Model for Astronomy and Astrophysics,
arXiv:2309.09122 (2023).

\bibitem{polak2024plotextract}
A. Polak, A. Morgan,
PlotExtract: A Chain-of-Thought Visual Reasoning Approach for Data Extraction from Scientific Plots,
arXiv:2404.08066 (2024).

\bibitem{hepdata}
E. Maguire et al.,
HEPData: a repository for high energy physics data,
Journal of Physics: Conference Series 898 (2017) 102006.

\bibitem{llmjudge}
L. Zheng et al.,
Judging LLM-as-a-Judge with MT-Bench and Chatbot Arena,
arXiv:2306.05685 (2023).

\bibitem{CMS_HSCP_2024}
CMS Collaboration,
Search for heavy long-lived charged particles with large ionization energy loss in proton-proton collisions at $\sqrt{s}=13$ TeV,
Physical Review D 111 (2025) 012011.


\bibitem{CMS_LLP_2024}
CMS Collaboration,
Search for light long-lived particles decaying to displaced jets in proton-proton collisions at $\sqrt{s}=13.6$ TeV,
Reports on Progress in Physics 88 (2025) 037801.


\bibitem{CMS_STOP_2025}
CMS Collaboration,
Search for top squarks in final states with many light-flavor jets and 0, 1, or 2 charged leptons in proton-proton collisions at $\sqrt{s}=13$ TeV,
Journal of High Energy Physics 10 (2025) 236.

\bibitem{lu2024ai_scientist}
Z. Lu et al.,
AI Scientist: Towards Fully Automated Scientific Discovery,
arXiv:2502.18864 (2024).

\bibitem{google_co_scientist}
Google Research,
Accelerating Scientific Breakthroughs with an AI Co-Scientist,
Google Research Blog (2024).

\bibitem{xu2023large_scale_review}
H. Xu et al.,
Large-Scale Multi-Agent Debate Improves Scientific Review,
arXiv:2311.16446 (2023).

\bibitem{duarte2024}
J. Duarte,
Novel Machine Learning Applications at the LHC,
Proceedings of ICHEP 2024, arXiv:2409.20413.


\end{thebibliography}
\end{document}